\documentclass[twocolumn, twocolappendix]{aastex631}

\usepackage{graphicx, graphics}
\usepackage{CJK}
\usepackage{bbm}
\usepackage{amsmath}
\usepackage{bm}
\usepackage{comment}
\usepackage{colortbl}
\usepackage{xcolor}
\usepackage[normalem]{ulem}
\definecolor{myblue}{HTML}{1f77b4} 

\newcommand{\arii}{[Ar\,{\scriptsize II}]}
\newcommand{\neii}{[Ne\,{\scriptsize II}]}

\begin{document}
\begin{CJK*}{UTF8}{gkai}

\title{The First JWST View of a 30-Myr-old Protoplanetary Disk Reveals a Late-stage Carbon-rich Phase}

\author[0000-0002-7607-719X]{Feng Long (龙凤)}
\altaffiliation{NASA Hubble Fellowship Program Sagan Fellow}
\affiliation{Lunar and Planetary Laboratory, University of Arizona, Tucson, AZ 85721, USA}
\email{fenglong@arizona.edu}

\author[0000-0001-7962-1683]{Ilaria Pascucci}
\affiliation{Lunar and Planetary Laboratory, University of Arizona, Tucson, AZ 85721, USA}

\author[0000-0001-8790-9011]{Adrien Houge}
\affiliation{Center for Star and Planet Formation, GLOBE Institute, University of Copenhagen, {\O}ster Voldgade 5-7, DK-1350 Copenhagen, Denmark}

\author[0000-0003-4335-0900]{Andrea Banzatti}
\affil{Department of Physics, Texas State University, 749 N Comanche Street, San Marcos, TX 78666, USA}

\author[0000-0001-7552-1562]{Klaus M. Pontoppidan}
\affiliation{Jet Propulsion Laboratory, California Institute of Technology, 4800 Oak Grove Drive, Pasadena, CA 91109, USA}

\author[0000-0002-5758-150X]{Joan Najita}
\affiliation{NSF's NOIRLab, 950 N. Cherry Avenue, Tucson, AZ 85719, USA}

\author[0000-0002-3291-6887]{Sebastiaan Krijt}
\affiliation{School of Physics and Astronomy, University of Exeter, Stocker Road, Exeter EX4 4QL, UK}

\author[0000-0001-8184-5547]{Chengyan Xie}
\affiliation{Lunar and Planetary Laboratory, University of Arizona, Tucson, AZ 85721, USA}

\author[0009-0008-8176-1974]{Joe Williams}
\affiliation{School of Physics and Astronomy, University of Exeter, Stocker Road, Exeter EX4 4QL, UK}

\author[0000-0002-7154-6065]{Gregory J. Herczeg (沈雷歌)}
\affiliation{Kavli Institute for Astronomy and Astrophysics, Peking University, Beijing 100871, China}
\affiliation{Department of Astronomy, Peking University, Beijing 100871, China}

\author[0000-0003-2253-2270]{Sean M. Andrews}
\affiliation{Center for Astrophysics \textbar\, Harvard \& Smithsonian, 60 Garden St., Cambridge, MA 02138, USA}

\author[0000-0003-4179-6394]{Edwin Bergin}
\affiliation{Department of Astronomy, University of Michigan, Ann Arbor, MI 48109, USA}

\author[0000-0003-0787-1610]{Geoffrey A. Blake}
\affiliation{Division of Geological \& Planetary Sciences, MC 150-21, California Institute of Technology, Pasadena, CA 91125, USA}

\author[0000-0002-5296-6232]{Mar\'{i}a Jos\'{e} Colmenares}
\affiliation{Department of Astronomy, University of Michigan, Ann Arbor, MI 48109, USA}

\author[0000-0001-6307-4195]{Daniel Harsono}
\affiliation{Institute of Astronomy, Department of Physics, National Tsing Hua University, Hsinchu, Taiwan}

\author[0000-0001-7152-9794]{Carlos E. Romero-Mirza}
\affiliation{Center for Astrophysics \textbar\, Harvard \& Smithsonian, 60 Garden St., Cambridge, MA 02138, USA}

\author[0000-0001-9222-4367]{Rixin Li (李日新)}
\altaffiliation{51 Pegasi b Fellow}
\affiliation{Department of Astronomy, Theoretical Astrophysics Center, and Center for Integrative Planetary Science, University of California
Berkeley, Berkeley, CA 94720-3411, USA}

\author[0000-0001-9352-0248]{Cicero X. Lu}
\affiliation{Gemini Observatory/NSF's NOIRLab, 670N. A'ohoku Place, Hilo, HI 96720, USA}

\author[0000-0001-8764-1780]{Paola Pinilla}
\affiliation{Mullard Space Science Laboratory, University College London, Holmbury St Mary, Dorking, Surrey RH5 6NT, UK.}

\author[0000-0003-1526-7587]{David J. Wilner}
\affiliation{Center for Astrophysics \textbar\, Harvard \& Smithsonian, 60 Garden St., Cambridge, MA 02138, USA}

\author[0000-0002-4147-3846]{Miguel Vioque}
\affiliation{European Southern Observatory, Karl-Schwarzschild-Str. 2, 85748, Garching bei Munchen, Germany}

\author[0000-0002-0661-7517]{Ke Zhang}
\affil{Department of Astronomy, University of Wisconsin-Madison, Madison, WI 53706, USA}

\author{the JDISCS collaboration}

\begin{abstract}
We present a JWST MIRI/MRS spectrum of the inner disk of WISE J044634.16$-$262756.1B (hereafter J0446B), an old ($\sim$34\,Myr) M4.5 star but with hints of ongoing accretion. The spectrum is molecule-rich and dominated by hydrocarbons. We detect 14 molecular species (H$_2$, CH$_3$, CH$_4$, C$_2$H$_2$, $^{13}$CCH$_2$, C$_2$H$_4$, C$_2$H$_6$, C$_3$H$_4$, C$_4$H$_2$, C$_6$H$_6$, HCN, HC$_3$N, CO$_2$ and $^{13}$CO$_2$) and 2 atomic lines (\neii\, and \arii), all observed for the first time in a disk at this age. The detection of spatially unresolved H$_2$ and Ne gas strongly supports that J0446B hosts a long-lived primordial disk, rather than a debris disk. The marginal H$_2$O detection and the high C$_2$H$_2$/CO$_2$ column density ratio indicate that the inner disk of J0446B has a very carbon-rich chemistry, with a gas-phase C/O ratio $\gtrsim$2, consistent with what have been found in most primordial disks around similarly low-mass stars. 
In the absence of significant outer disk dust substructures, inner disks are expected to first become water-rich due to the rapid inward drift of icy pebbles, and evolve into carbon-rich as outer disk gas flows inward on longer timescales. The faint millimeter emission in such low-mass star disks implies that they may have depleted their outer icy pebble reservoir early and already passed the water-rich phase. 
Models with pebble drift and volatile transport suggest that maintaining a carbon-rich chemistry for tens of Myr likely requires a slowly evolving disk with $\alpha-$viscosity $\lesssim10^{-4}$. 
This study represents the first detailed characterization of disk gas at $\sim$30\,Myr, strongly motivating further studies into the final stages of disk evolution. 

\end{abstract}


\section{Introduction} \label{sec:intro}
Gas in protoplanetary disks has a major impact on the formation and evolution of planetary systems. The lifetime of the gas disk directly constrains the timescale of giant planet formation. The presence of disk gas can significantly alter system architectures by driving planet migration and reshaping orbital configurations (see e.g., \citealt{Paardekooper2023}).
Additionally, the atmospheric compositions and potential habitability of exoplanets are closely linked to the disk gas they accrete. Observational and theoretical efforts have also been made to link the C/O ratio of exoplanet atmospheres to disk chemistry models to reveal the planet's formation location and dynamic history (e.g., \citealt{Oberg2011, Madhusudhan2012, Molliere2022}). 
Obtaining observational constraints of the gas disk evolution and its chemical composition across disk radii is thus essential to develop a complete model of planet formation.

Spitzer/IRS spectra with moderate resolution ($R\sim600$) have revealed rich volatile chemistry within the inner few AU of young protoplanetary disks and identified a series of water, OH, CO$_2$, C$_2$H$_2$, and HCN emission lines at $\sim$5--38\,$\mu$m (e.g., \citealt{Carr2008Sci, Carr2011, Pontoppidan2010ApJ...720..887P, Salyk2011}). Though faint disks around mid-to-late M stars (with spectral type later than M3) were rarely targeted with Spitzer, they were found to be different from their solar analogs, showing brighter C$_2$H$_2$ emission over HCN \citep{Pascucci2009} and weaker H$_2$O lines \citep{Pascucci2013}. Comparison of the derived gas properties to thermo-chemical models (e.g., \citealt{Najita2011_model}) has thus suggested enhanced carbon chemistry with C/O ratios of $\sim1$ in their disks \citep{Pascucci2013}.

The Medium Resolution Spectrometer (MRS; \citealt{Wells2015}) on board of JWST/MIRI (the Mid-Infrared Instrument, \citealt{Rieke2015}) now offers significantly improved sensitivity and spectral resolution ($R\sim2000-4000$, see Table 3 in \citealt{Pontoppidan2024}), presenting new opportunities to investigate chemical variations in disks around different stellar types. 
Recent JWST observations of known C-rich disks around very low-mass stars, such as 2MASS J16053215$-$1933159 (hereafter as J160532, M5, \citealt{Tabone2023NatAs...7..805T}) and ChaI-147 (M5.5, \citealt{Arabhavi2024Sci...384.1086A}), have further identified a large number of hydrocarbon molecules (see Table~\ref{tab:mdwarf-comp}), including the first detections of C$_6$H$_6$ (benzene) in protoplanetary disks. In contrast, the MIRI/MRS spectrum of Sz\,114, a star with a similar spectral type as the two above, is dominated by water emission \citep{Xie2023}. This suggests that factors beyond host star properties may play important roles in disk chemical evolution. Theoretical models by \citet{Mah2023} indicate that, under the combined effects of icy pebble drift and gas accretion, the inner disk would initially experience a drop in gas C/O ratio due to ice sublimation, then followed by an increase from the outer gas inflow. The transition point is expected to occur earlier in disks around lower-mass stars due to their closer-in ice lines and shorter viscous timescales. Additionally, the properties of dust traps could substantially influence the chemical evolution pathways (e.g., \citealt{Kalyaan2023, Mah2024}).

Although typical disk lifetimes are known to be only a few Myrs (e.g., \citealt{Hernandez2008, Ribas2014}), recent studies have identified a number of accreting disks surrounding very low-mass stars with ages of 30--50\,Myr (e.g., \citealt{Boucher2016, Murphy2018, Silverberg2020}). This old sample thus provides a unique prospect to enable gas evolution studies across tens of Myrs. Here we present JWST/MIRI observations for one such old-accreting disk around the M4.5 star WISE J044634.16$-$262756.1B (hereafter J0446B; Gaia DR3 coordinate: 04:46:34.25 $-$26:27:55.57, \citealt{gaiadr3}). This represents the first study of gas-rich disks at $\sim$30\,Myr old, including the first detections of H$_2$, \neii\,, and large numbers of hydrocarbons at such old ages. The paper is structured as follows. In Section~\ref{sec:obs}, we describe our target and the JWST observations. The resulting spectrum and slab model fits for identified lines are presented in Section~\ref{sec:results}. We then discuss the implications of these line detections for disk chemical evolution and planet formation in Section~\ref{sec:diss}, and summarize the key points in Section~\ref{sec:summ}.

\section{The Target and Observations} \label{sec:obs}

\begin{figure*}[!t]
\centering
    \includegraphics[width=0.9\textwidth]{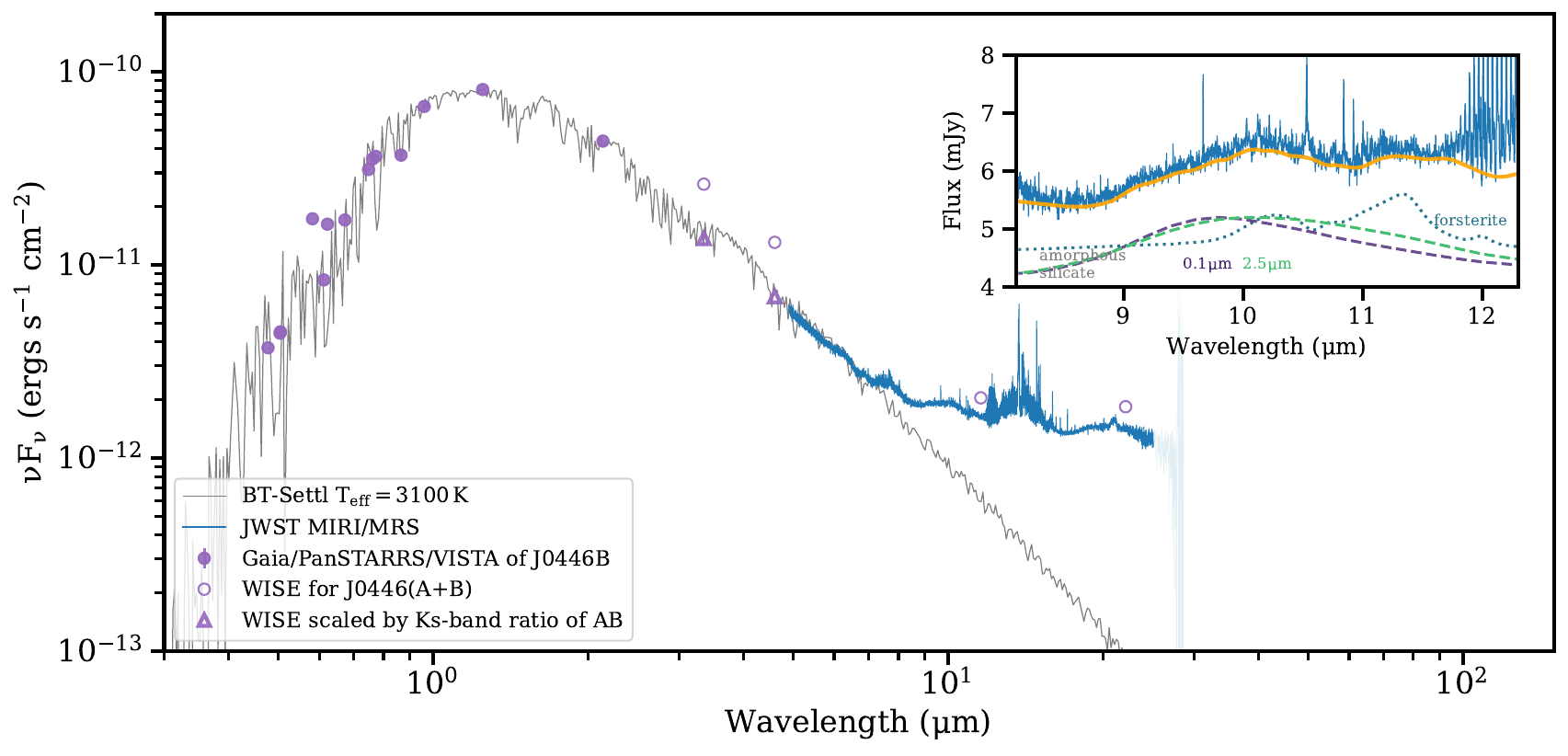}
\caption{The spectral energy distribution of J0446B including the JWST MIRI/MRS spectrum in blue (the noisy long-wavelength range is marked with lighter color). The grey curve shows a stellar photospheric model with $T_{\rm eff}=3100\,K$ \citep{Allard2012}. The four WISE band photometry encompassed both stellar components in the binary system of J0446; the triangles represent the contribution of J0446B to W1 and W2 bands assuming the same flux ratio as the Ks band. The insert panel highlights the 10\,$\mu$m silicate features, where the orange curve marks the identified continuum, two dashed lines represent emissivity curves of amorphous silicates (olivine) for a single grain size of 0.1 and 2.5\,$\mu$m \citep{Jaeger1994}, and the dotted line represents that for crystalline silicate of forsterite \citep{Koike2003}.  \label{fig:sed}}
\end{figure*}


\subsection{J0446B: an old disk with hints of accretion} \label{sec:target}
WISE J044634.16$-$262756.1 was identified as a source with strong infrared excess through the Disk Detective citizen science project, with a high probability of membership in the $42^{+6}_{-4}$\,Myr-old Columba association \citep{Silverberg2020}. Images from Pan-STARRS and Gaia revealed it as consisting of two stars (J0446A - the SW component, and J0446B - the NE component) separated by 2$\farcs$3, corresponding to $\sim$189\,AU at a distance of 82\,pc \citep{gaiadr3}. Recently, \citet{Luhman2024} reassessed the memberships and ages of nearby young moving groups based on Gaia DR3 and assigned J0446A and J0446B to $\chi^{1}$ For, a region physically related to Columba with an age of $33.7^{+2.0}_{-1.9}$\,Myr. In this paper, we adopt this new age determination based on lithium depletion models. The absence of lithium in the Gemini/GMOS spectra of J0446A and J0446B also aligns with such an old age \citep{Silverberg2020}. 
Evidence of accretion was suggested in both disks by \citet{Silverberg2020}, though the $H_{\alpha}$ line widths are at the borderline of separating accretion from chromospheric activity. J0446B shows stronger and broader $H_{\alpha}$ line emission than J0446A, for the latter our photospheric-like MIRI spectra (see Appendix~\ref{sec:J0446A}) is more in agreement with $H_{\alpha}$ arising from chromospheric activity. In this paper, we thus focus on the gas-rich disk around J0446B, which has an estimated mass accretion rate of $2.5\times10^{-11} M_{\odot}/yr$ \citep{Silverberg2020}. This rate falls at the lower end of the reported range for young 0.1--0.2\,$M_{\odot}$ stars of 1--10\,Myr (see \citealt{Manara2023} for a compilation and references therein).

The M6 spectral type (2800\,K) measured by \citet{Silverberg2020} results in optical emission that is much fainter than observed (for both stars). We determined a spectral type of M4.5 by comparing the Gaia XP spectra of J0446B \citep{gaiaspec} to XP spectra of objects in the TW Hya Association (as described in \citealt{Luhman2023}, see Figure~\ref{fig:sed_A} in the Appendix). This corresponds to an effective temperature of $\sim$3100\,K \citep{Pecaut2013}, which provides a much-improved fit to the broadband SED. The stellar luminosity of J0446B is 0.016\,$L_{\odot}$, measured from the VISTA J-band magnitude of 11.828, the bolometric correction from \citet{Pecaut2013}, and zero-point flux of 3.013$\times$10$^{35}$ erg\,s$^{-1}$. Based on \citet{Somers2020} models, we estimate the stellar mass to be 0.13--0.22\,$M_{\odot}$, depending on whether age is fixed or not.
The dust disk of J0446B was detected but unresolved with ALMA observations (with beam size of $\sim$0\farcs6), yielding a total flux of 1.2\,mJy at 0.9\,mm (K. Flaherty, private communication), which corresponds to a dust mass of only 0.1\,$M_{\earth}$, assuming optically thin emission and adopting the same opacity as used in \citet{Pascucci2016} and a dust temperature of 20\,K. This is about one order of magnitude lower than the typical dust mass in disks around young stars of similar types \citep{Manara2023}.

\subsection{JWST observations and data reduction} \label{sec:jwst-obs}
J0446B was observed with MIRI/MRS on 2024 January 29, as part of the JWST Cycle 2 program GO-3153 (PI. F.~Long), which was designed to reveal the nature of this class of old accreting disks. The observation started with the target acquisition procedure using a neutral density filter, which placed the brighter IR source (i.e. J0446B in the binary system) in the center of the field of view. The four-point dither pattern (optimized for point sources) was used for thermal background subtraction. To ensure high S/N $\sim$100 at the longest wavelengths without saturation, each sub-band was integrated with 60 groups per ramp in the FASTR1 mode with a total integration time of $\sim$16\,min. All three sub-bands were selected to cover the full wavelength range from 4.9 to 28.6\,$\mu$m.

As part of the JWST Disk Infrared Spectral Chemistry Survey (the JDISCS collaboration), our data reduction follows the procedure established in \citet{Pontoppidan2024}. 
Briefly, individual cubes were built for each exposure, channel, and sub-band using \texttt{callwebb\_step2} of the JWST Calibration Pipeline version 1.15.0 and Calibration Reference Data System context 1254. After background subtraction using the two opposite dither positions, a 1D spectrum was extracted with an aperture that increased linearly with wavelength within each sub-band. We note that J0446A and J0446B are well separated in every wavelength channel, thus the extracted spectrum is not contaminated by the companion, nor is it affected by the background subtraction. Lastly, the relative spectral response functions derived from observations of asteroids (GO-1549 and GO-3034) were applied to our data spectrum to remove remaining fringe patterns and provide accurate absolute spectro-photometric calibration.

\begin{figure*}[!t]
\centering
    \includegraphics[width=0.9\textwidth]{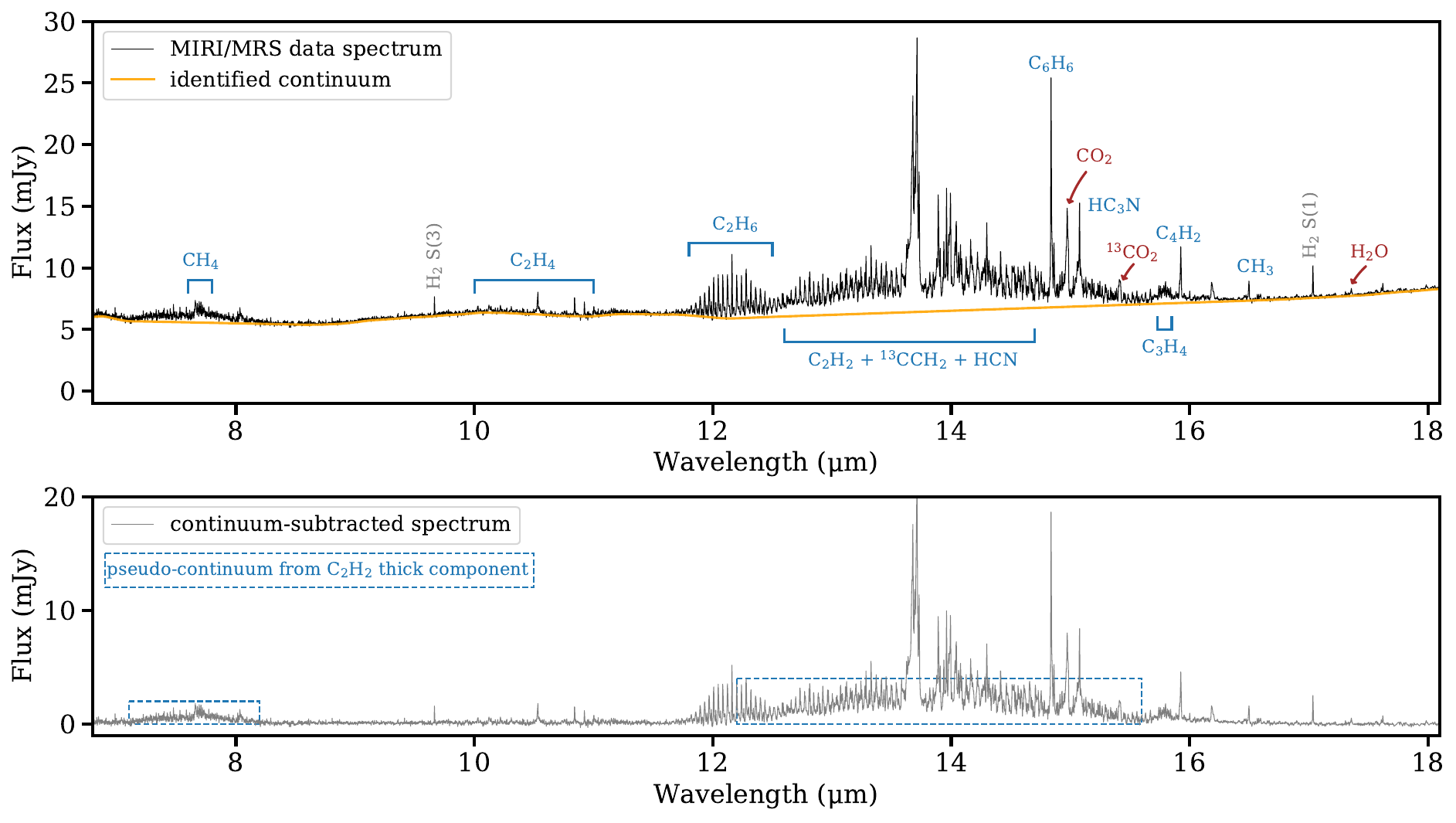}
\caption{{\bf\textit{Top:}} The MIRI-MRS spectrum (black) of the disk of J0446B and the estimated continuum emission (orange). The shorter wavelength affected by stellar absorption and the wavelength range longer than 18\,$\mu$m with worse data quality and no lines-of-interest are cut out. The identified molecules are marked out.  {\bf\textit{Bottom:}} The continuum-subtracted spectrum (grey), where the regions with pseudo-continuum emission from the optically thick C$_2$H$_2$ component are highlighted. \label{fig:overview}}
\end{figure*}

\section{Analysis and Results} \label{sec:results}

\subsection{Spectrum Overview and Continuum Subtraction}
The full spectral energy distribution of J0446B, including the JWST MIRI/MRS spectrum, is shown in Figure~\ref{fig:sed}. The wavelength range short of $\sim7\,\mu$m in our MIRI spectrum well aligns with the stellar photosphere model of 3100\,$K$, suggesting that the region very close to the central star is depleted of dust materials (see review of \citealt{Espaillat2014}). Within this wavelength range, we also see clear stellar absorption features of CO and H$_2$O (see Figure~\ref{fig:stellar} in the Appendix). 
At longer wavelengths, both the 10 and 20\,$\mu$m silicate features are visible. Additionally, we find two other bumps around 7.5 and 14\,$\mu$m, which are likely due to a pseudo-continuum produced by optically thick molecular emission \citep{Tabone2023NatAs...7..805T}.

The MIRI/MRS spectrum of J0446B shows rich molecular line emission, particularly within the wavelength range of 12--16\,$\mu$m (Figure~\ref{fig:overview}). Initial inspection of the spectrum based on the HITRAN database \citep{Gordon2022} and the \texttt{iSLAT} tool \citep{Jellison2024} indicates that the inner disk of J0446B contains a large number of hydrocarbon molecules, including CH$_4$ (peaking at 7.65\,$\mu$m), C$_2$H$_4$ (10.53\,$\mu$m), C$_2$H$_6$ (12.17\,$\mu$m), C$_2$H$_2$ (13.69\,$\mu$m), $^{13}$CCH$_2$ (13.73\,$\mu$m), and C$_4$H$_2$ (15.92\,$\mu$m). The bright emission at 14.85\,$\mu$m corresponds to the Q-branch of the hot bending mode $\nu$4 of Benzene, C$_6$H$_6$, as described in \citet{Tabone2023NatAs...7..805T}, which has line peak flux comparable to C$_2$H$_2$. Following the recent study of a carbon-rich disk around ChaI-147 \citep{Arabhavi2024Sci...384.1086A}, we also find emission lines of C$_3$H$_4$ and CH$_3$ around the corresponding wavelengths of 15.80 and 16.48\,$\mu$m, respectively. Two nitrogen-bearing molecules, HCN and HC$_3$N (15.08\,$\mu$m) are clearly detected, and the HCN emission is heavily blended with C$_2$H$_2$. 
CO$_2$ (14.98 and 16.18\,$\mu$m), along with its isotopologue $^{13}$CO$_2$ (15.41\,$\mu$m), is the only robustly identified oxygen-bearing molecule in the spectrum of J0446B. The H$_2$O emission, though widely detected in T Tauri disks (see review of \citealt{Pontoppidan2014}), is very weak and marginally detected in this disk. Multiple molecular hydrogen (H$_2$) lines and two atomic lines (\neii\, and \arii) are also identified and will be discussed in Sec.~\ref{sec:h2}.

To facilitate further analysis of the molecular lines, the dust continuum emission needs to be removed from the observed spectrum. Following the procedure outlined in \citet{Pontoppidan2024}, we first computed the underlying continuum using an iterative median filter applied over five rounds, with a window of 65 and $\sim$95 wavelength channels for long and short wavelengths, respectively. We excluded wavelength ranges of 7.1--8.5 and 12.0--16.5\,$\mu$m, where optically thick C$_2$H$_2$ emission could produce a pseudo-continuum. At wavelengths $>$16.5\,$\mu$m, we used line-free regions identified by \citet{Banzatti2024arXiv} to apply a small wavelength-dependent offset to better match the expected continuum level for a more robust constraint on the water emission there. Such offset is difficult to identify in other wavelength ranges of this disk due to its rich organic emissions. The final continuum was then smoothed with a second-order Savitzky-Golay filter. 
Both the identified continuum and the continuum-subtracted spectrum are shown in Figure~\ref{fig:overview}. Similar to the disk around M6 star J160532 in the 5--10\,Myr-old Upper Sco association \citep{Tabone2023NatAs...7..805T}, two emission bumps at 7.05--8.00 and 12.0--16.0\,$\mu$m remain after the continuum removal, indicating the presence of an optically thick component of C$_2$H$_2$.

The 10\,$\mu$m silicate feature is commonly used to assess the evolution of dust grains in protoplanetary disks, as the shape and strength of the solid-state features are largely determined by dust grain size and composition \citep{Kessler-Silacci2006, Furlan2011}. As highlighted in Figure~\ref{fig:sed} (insert plot), the 10\,$\mu$m feature in the disk of J0446B is rather broad and flat, with a peak-over-continuum flux of 1.12, where the comparing continuum is estimated based on fluxes around 8.8 and 12.2\,$\mu$m. 
Compared to emissivity curves of different grain sizes \citep{Jaeger1994}, the peak location in the J0446B spectrum suggests that silicate grains with a few micron sizes are present in the disk, consistent with earlier studies of disks around very low mass objects \citep{Apai2005Sci, Pascucci2009}. 
The spectrum also likely includes a crystalline feature of forsterite at 11.3\,$\mu$m. A detailed dust composition analysis will be presented in a future paper.

\subsection{Molecular Line Modeling}

\begin{figure*}[!th]
\centering
    \includegraphics[width=\textwidth, trim=50 0 50 20, clip]{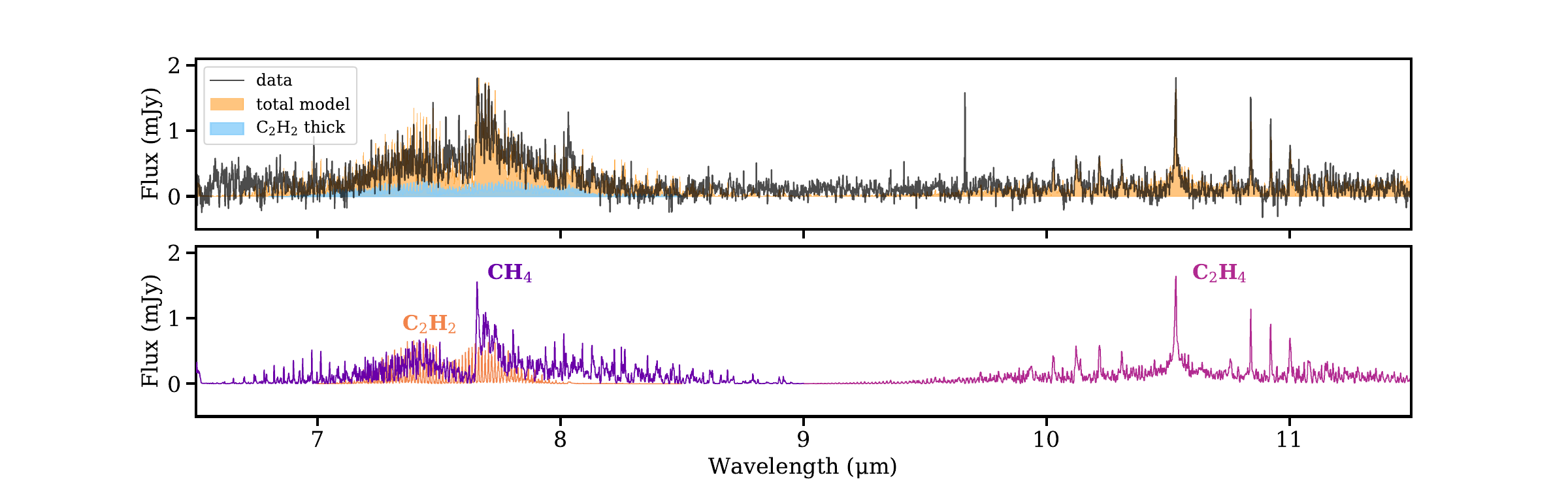}
    \includegraphics[width=\textwidth, trim=50 0 50 20, clip]{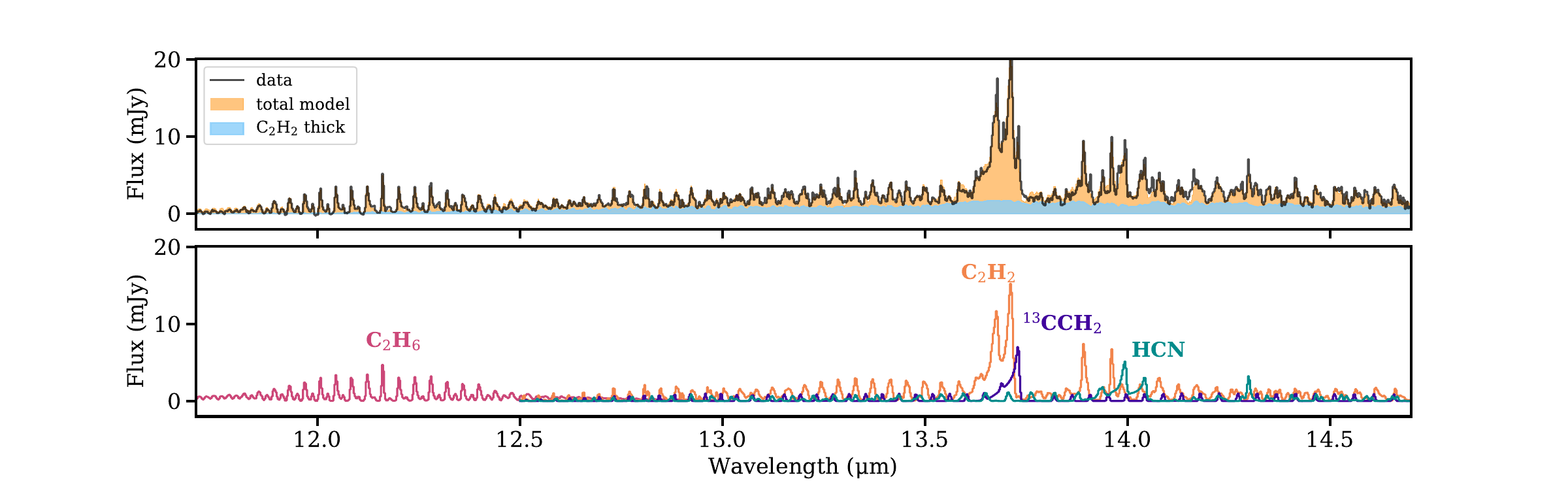}
    \includegraphics[width=\textwidth, trim=50 0 50 20, clip]{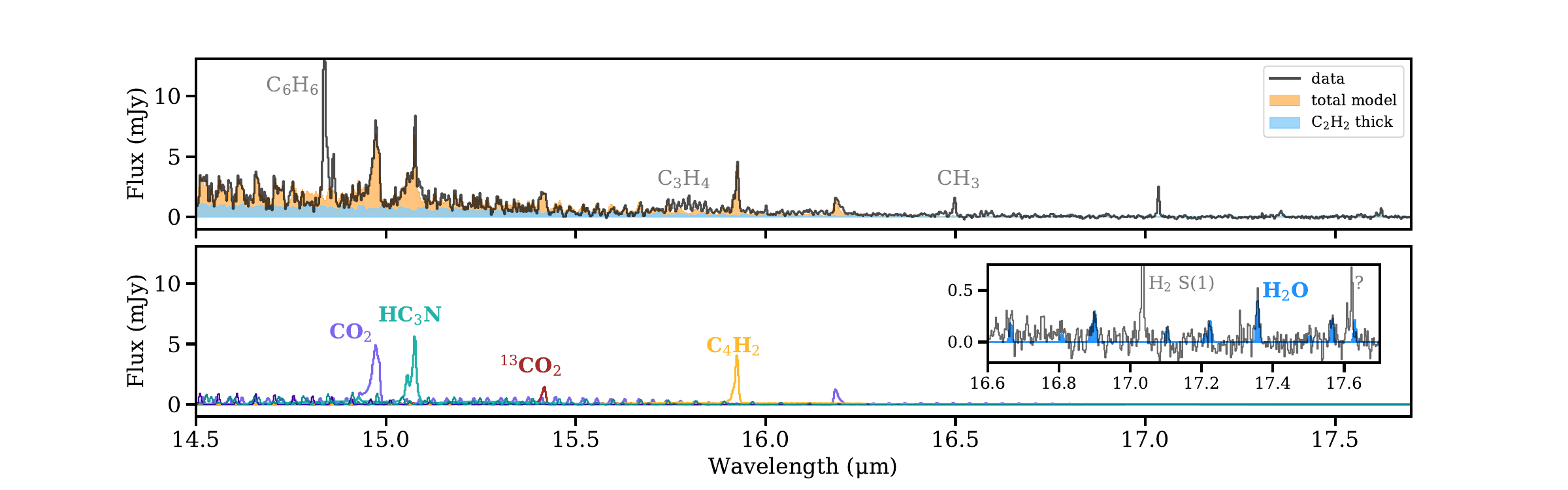}
\caption{The comparison of data and model spectra. {\bf \textit{Top panels:}} The continuum-subtracted data spectrum (black) with a model spectrum (orange-shaded region) composed of all fitted species. The optically thick emission of C$_2$H$_2$ is highlighted in light blue. {\bf \textit{Bottom panels:}} Model spectra for individual molecules in each wavelength segment. The insert plot highlights the marginal detection of H$_2$O, and the possible detection of emission lines around 17.62\,$\mu$m remains unidentified. \label{fig:slab-model} }
\end{figure*}

Physical conditions of the inner disk regions can be explored through the rich molecular emissions at mid-infrared wavelengths. To characterize the gas properties, we employ the plane-parallel slab model that describes line emission with three free parameters for a particular molecule: column density $N$, gas temperature $T$, and emitting area $A$, with the latter often expressed as the effective emitting radius $R_{\rm slab}$ with $A$=$\pi$$R{_{\rm slab}}^{2}$. While the slab models only represent average conditions in the disk atmosphere due to the complexity of the disk environment, these simple models have been successfully applied in previous studies and can well reproduce molecular spectra observed with Spitzer and JWST (e.g., \citealt{Carr2008Sci, Carr2011, Salyk2011, Grant2023, Schwarz2024}).  

In this work, we adopt the slab modeling Python package \texttt{iris} \citep{Munoz-Romero2023zndo} with spectroscopic data from the HITRAN database \citep{Gordon2022}, which was recently applied to the study of the MIRI-MRS spectrum of AS\,209 by \citet{Munoz-Romero2024}. The model assumes local thermal equilibrium (LTE) for level populations but includes a treatment of optical depth effects and line overlap, following the procedure outlined in \citet{Tabone2023NatAs...7..805T}. The latter implementation is especially important for molecules with closely spaced lines and at high column densities, where line overlap can result in an effectively optically thick continuum over a wide wavelength range. This situation applies directly to our case. 
The line shape is assumed to be Gaussian with a width of $\sigma=2$\,km\,s$^{-1}$. This choice is consistent with previous studies\footnote{Our strategy is the same as \citet{Tabone2023NatAs...7..805T}, while \citet{Arabhavi2024Sci...384.1086A} assumed $\sigma=2$\,km\,s$^{-1}$ for the turbulent broadening alone as thermal broadening is typically small in disks around low-mass stars.}, to account for the unknown contribution of turbulent broadening, and allows direct comparisons with literature results. For optically thick lines, the column densities are expected to roughly scale inversely with the line width. The emission models are first generated at a sufficiently high resolution ($R\sim10^5$) to account for overlapping lines and then resampled to the data wavelength grid. 
The model fitting is performed with the Bayesian Nested Sampling Python package \texttt{dynesty} \citep{Speagle2020}, in which the bounding distribution of multiple ellipsoids and random walk sampling are adopted. The stopping criterion is set by the change in the remaining evidence ($\mathcal{Z}$, marginal likelihood) when $\Delta \log \mathcal{Z} \leq 0.001 (n_{live} - 1) + 0.01 $, where $n_{live}$  = $10\times n_{dim}$ with $n_{dim}$ as the number of parameters in the model.

We started by fitting the prominent C$_2$H$_2$ emission region, considering both optically thin and thick components with independent emitting area and excitation temperature, similar to \citet{Tabone2023NatAs...7..805T}. Given the wide wavelength contribution of the optically thick C$_2$H$_2$ emission that overlaps with emissions from many other molecules, we included in the fit also C$_2$H$_6$, $^{13}$CCH$_2$, HCN, CO$_2$, and HC$_3$N. The other molecules were fitted with the same emitting area and temperature as the C$_2$H$_2$ optically thin component to speed up the convergence\footnote{Without the simultaneous fit of CO$_2$ and HC$_3$N at the longer wavelength, the retrieved optically thick component tends to significantly overproduce emission around 15.5\,$\mu$m.}. We restricted the fitting within the wavelength range of 11.8--15.65\,$\mu$m but excluded the range of 14.81--14.85\,$\mu$m where C$_6$H$_6$ emission dominates and lacks spectroscopic data in the HITRAN database. Uniform priors were adopted for emitting area and column density, with $\log_{10}A = \mathcal{U} (-4, 2) \text{ au}^{2}$ and $\log_{10} N = \mathcal{U} (14, 20)/(20, 24) \text{ cm}^{-2}$ for thin(others)/thick components.  
The excitation temperature used a Normal prior which centered around 400\,K with a standard deviation of 200\,K. 
Our test model showed that the choice of prior distribution does not affect the fitting results. In this work, we choose the median of the posterior distribution as the best-fit value for each parameter, as listed in Table~\ref{tab:slab-model} along with the 3$\sigma$ uncertainties. 

We inspected the goodness of the fit by plotting the best-fit model on top of the data spectrum. The $^{12}$CO$_2$ model with the same emitting area and temperature as others failed to reproduce its hot-band Q-branch at 16.2\,$\mu$m, likely indicating an underestimation of column density as demonstrated by \citet{Grant2023}. We thus proceeded by subtracting the optically thick model of C$_2$H$_2$ and fitting again individual molecules with varying $A$, $N$, and $T$. Considering still the line blending of C$_2$H$_2$, HCN, and $^{12}$CO$_2$, we first fitted all three molecules with their isotopologues ($^{13}$CCH$_2$ and $^{13}$CO$_2$) together across the wavelength range of 12.5--16.4\,$\mu$m. The resulting column density for $^{12}$CO$_2$ is about an order of magnitude higher than the previous model and well matches multiple $^{12}$CO$_2$ emission features. We therefore adopted the new model for the five species as summarized in Table~\ref{tab:slab-model}. After subtracting the best-fit models for the above species across the full data wavelength range, we then fitted CH$_4$, C$_2$H$_4$, C$_2$H$_6$, and C$_4$H$_2$ within specific wavelength regions (see Table~\ref{tab:slab-model}) for each molecule. The comparison of the data and best-fit model spectra is shown in Figure~\ref{fig:slab-model}. The mismatch around 7.5\,$\mu$m is likely related to the stellar photospheric absorption.

We retrieved very high column density ($\sim10^{22}$ cm$^{-2}$) for the C$_2$H$_2$ thick component, which has a similar excitation temperature to the thin component but with a smaller emitting area.
As a reference, the typical column density for C$_2$H$_2$ in T Tauri disks is within the range of $<10^{13}-10^{16}$\,cm$^{-2}$ \citep{Salyk2011}.
Other hydrocarbon molecules (except for C$_4$H$_2$) share similar excitation conditions to the C$_2$H$_2$ thin component, with $T\sim250-300$\,K and $R_{\rm slab}\sim0.05-0.1$\,AU. The more extended emitting area for C$_4$H$_2$ is, however, consistent with chemical model predictions \citep{Kanwar2024arXiv}.
HCN appears to be co-spatial with C$_2$H$_2$, while CO$_2$ might reside either further out or in a deeper layer with a slightly lower $T$ of $\sim$214\,K (and $R_{\rm slab}\sim0.1$\,AU).

Water emission in the disk of J0446B is very weak and marginally detected (see the inset plot in Figure~\ref{fig:slab-model} for the 17\,$\mu$m range). Adopting the same excitation conditions to the thin component of C$_2$H$_2$, we estimated an upper limit for the water column density of $\sim10^{18}$ cm$^{-2}$, consistent with reported values for water detections in T Tauri disks (e.g., \citealt{Grant2023, Schwarz2024, Munoz-Romero2024}). 
However, the total emitting water molecule number ($N \pi R^2$) of 3.5$\times10^{42}$ (corresponding to a mass of $1.8\times10^{-8}\,M_{\earth}$) is significantly lower than that in T Tauri disks.

\begin{deluxetable}{lcccc}[!th]
\tabletypesize{\scriptsize}
\tablecaption{Slab Model Results\label{tab:slab-model}}
\tablewidth{0pt}
\tablehead{
\colhead{Species}  &  \colhead{Fitting Range} & \colhead{$R_{\rm slab}$} & \colhead{$\log N$} & \colhead{$T$}  \\ 
\colhead{} &  \colhead{($\mu$m)} & \colhead{(AU)} & \colhead{(cm$^{-2}$)} & \colhead{(K)}
} 
\startdata
CH$_4$ &  7--8.5 & 0.034$\pm$0.006 & 20.54$\pm$0.21 & 309$\pm$21 \\
C$_2$H$_4$ &  9.7--11.5 & 0.065$\pm$0.013 & 17.75$\pm$0.11 & 240$\pm$15 \\
C$_2$H$_6$ &  11.8--12.5 & 0.195$\pm$0.025 & 18.45$\pm$0.13 & 200$\pm$9 \\
C$_{2}$H$_{2}$ - thick &  11.8--15.65 & 0.019$\pm$0.002 & 22.54$\pm$0.26 & 288$\pm$13 \\
C$_{2}$H$_{2}$ - thin &  12.6--16.3 & 0.072$\pm$0.002 & 18.36$\pm$0.05 & 295$\pm$10 \\
$^{13}$CCH$_{2}$   & 12.6--16.3 & 0.103$\pm$0.014 & 17.45$\pm$0.18 & 263$\pm$24 \\
C$_4$H$_2$\tablenotemark{\footnotesize a}  & 15.85--16.0 &  0.48$^{+3.38}_{-0.34}$ & 14.89$^{+1.31}_{-1.89}$ & 195$\pm$20 \\
\hline
CO$_2$ &  12.6--16.3 & 0.097$\pm$0.013 & 18.46$\pm$0.17 & 214$\pm$15 \\
$^{13}$CO$_2$\tablenotemark{\footnotesize a}  & 12.6--16.3 & 1.10$^{+1.48}_{-0.84}$ & 14.54$^{+1.23}_{-0.54}$ & 170$\pm$23 \\
\hline
HCN &  12.6--16.3 & 0.056$\pm$0.005 & 18.45$\pm$0.19 & 302$\pm$21 \\
HC$_3$N\tablenotemark{\footnotesize b} & 11.8--15.65 & 0.085$\pm$0.002 & 16.33$\pm$0.01 & 263$\pm$6 \\
\enddata
\tablecomments{The listed uncertainties correspond to statistical uncertainties and are likely underestimated. The spectroscopic data for $^{13}$CCH$_2$ in the HITRAN database is incomplete so the derived column density should be taken cautiously.}
\tablenotetext{a}{The emitting area and column density for C$_4$H$_2$ and $^{13}$CO$_2$ are not well constrained. }
\tablenotetext{b}{The model for HC$_3$N is taken from the initial combined fit to constrain the optically thick component of C$_{2}$H$_{2}$.}
\end{deluxetable}

\subsection{$H_2$ emission and Atomic lines} \label{sec:h2}
\begin{figure*}[!t]
\centering
    \includegraphics[width=0.62\textwidth]{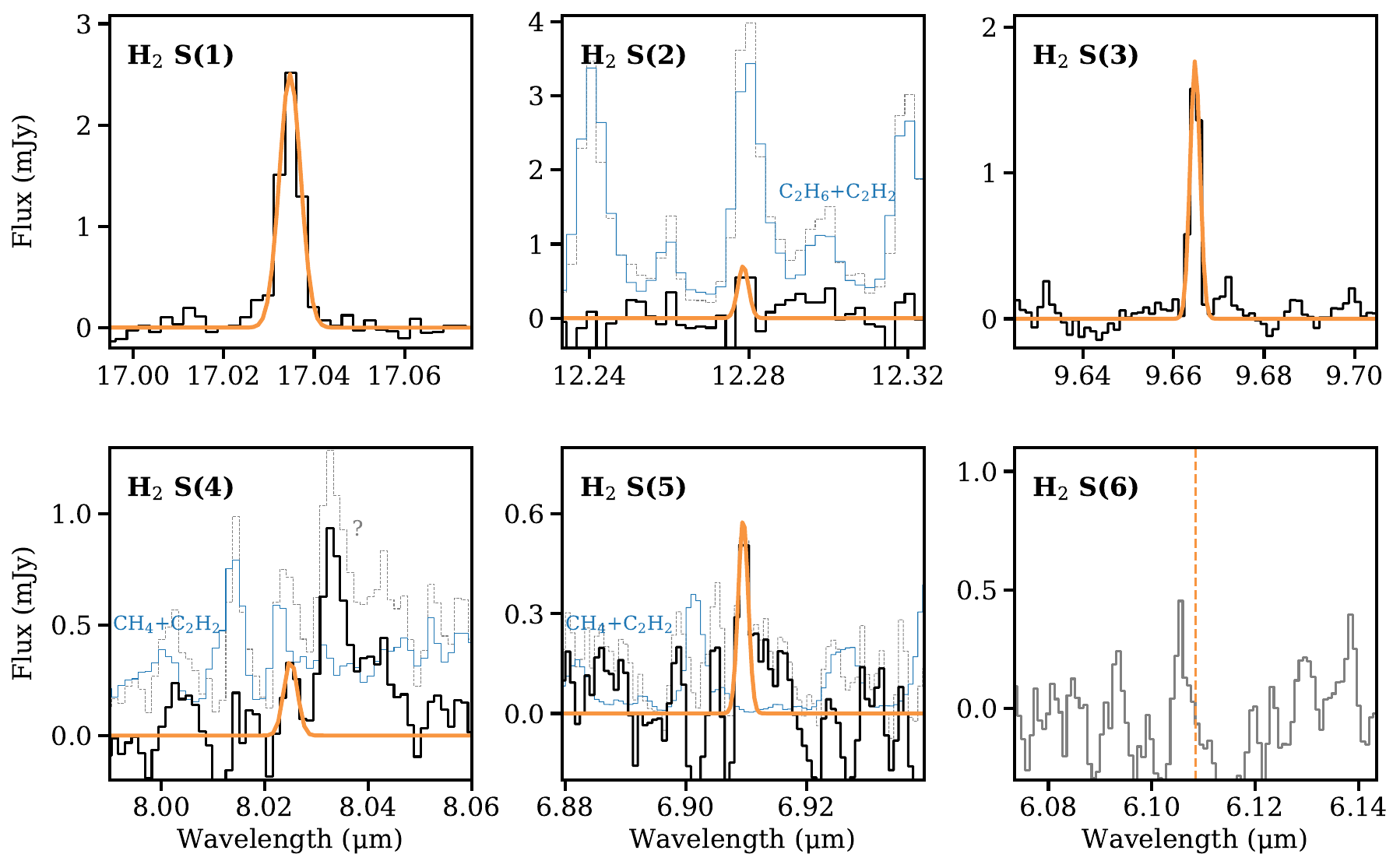}
    \includegraphics[width=0.33\textwidth]{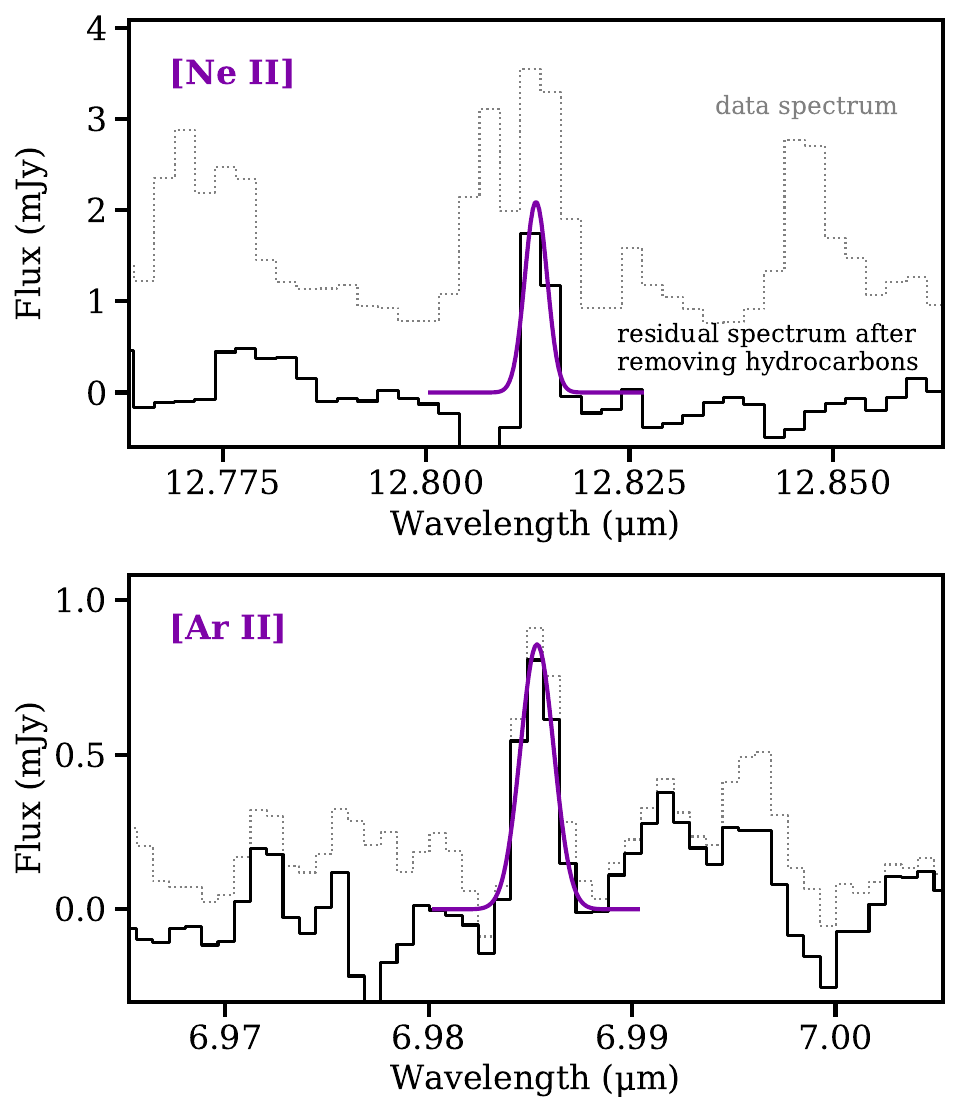}
\caption{{\bf \textit{Left: }}Zoomed-in spectra around molecular hydrogen pure rotational lines. Fitted Gaussian profiles are overplotted for detected lines. H$_2$ S(2), S(4), S(5) are blended with hydrocarbon emissions, for which the slab models are indicated in blue with the original data spectra shown in grey. The vertical dashed lines mark the line center for non-detection. {\bf \textit{Right: }} The detection of \neii\, at 12.814\,$\mu$m and \arii\, at 6.985\,$\mu$m.  \label{fig:h2}}
\end{figure*}

\begin{deluxetable}{lcc}
\tabletypesize{\scriptsize}
\tablecaption{Line Fluxes\label{tab:line-fluxes}}
\tablewidth{0pt}
\tablehead{
\colhead{Species}  & \colhead{Wavelength} & \colhead{Fluxes}   \\ 
\colhead{} & \colhead{($\mu$m)} & \colhead{($\rm 10^{-16}\,erg\,s^{-1}\,cm^2$)} \\
} 
\colnumbers
\startdata
H2 (0,0) S(1) & 17.035 & 1.52 \\
H2 (0,0) S(2) & 12.279 & 0.47 \\
H2 (0,0) S(3) & 9.665 & 1.55 \\
H2 (0,0) S(4) & 8.025 & 0.52 \\
H2 (0,0) S(5) & 6.909 & 0.76 \\
\hline
$[\rm Ne\,II]$ & 12.814 & 1.27 \\
$[\rm Ar\,II]$ & 6.985 & 1.08 \\ 
\hline
C$_2$H$_2$ & 13.65--13.72 & 105.5 \\
H$_2$O & 17.19--17.25 & $<$0.82\tablenotemark{\footnotesize a}\\
\enddata
\tablenotetext{a}{The reported upper limit is 3$\times$ the flux from the water model spectrum as shown in Figure~\ref{fig:slab-model}, thus providing a stringent lower limit of the line ratios discussed in Section~\ref{sec:diss}. }
\end{deluxetable}

The wide wavelength range of MIRI/MRS covers multiple H$_2$ emission lines, among which five pure rotational $\nu=0-0$ transitions, from S(1) to S(5), are detected in the residual spectra after the removal of the above models (see Figure~\ref{fig:h2}). These H$_2$ emission lines have been previously identified in a few disks around very low-mass young stars ($\sim$1--10\,Myr, \citealt{Pascucci2013, Xie2023, Franceschi2024}). 
We calculated the line fluxes through a fitted Gaussian profile centered at the corresponding line wavelength and summarized in Table~\ref{tab:line-fluxes}. Due to the significant blending of the H$_2$ S(2) and S(4) with hydrocarbon emissions, and the weak detection of H$_2$ S(5), we used the two isolated strong lines H$_2$ S(1) and S(3) to constrain the gas properties (e.g., \citealt{Thi2001}, \citealt{Lahuis2007}). 
Using the line ratio of these transitions, we calculated a temperature of 369$\pm1$\,K for the H$_2$ gas, within the reasonable range of inner surface temperatures of M-star disks \citep{Walsh2015, Kanwar2024}. Given this slightly warmer temperature, the emitting H$_2$ gas may be located closer to the star than most hydrocarbons. 
A visual inspection of the data cube did not reveal any extended H$_2$ emission, suggesting that the observed emission is likely originating from the disk. Although the possibility of tracing a small-scale unresolved disk wind can not be completely ruled out, given the moderate spatial resolution of MIRI/MRS (0$\farcs$2-0$\farcs$3). 
Assuming optically thin emission, we obtained a lower limit for the gas mass within the inner few AU to be only 0.01\,$M_{\earth}$.

Two atomic lines are also detected: \arii\, at 6.985\,$\mu$m and \neii\, at 12.814\,$\mu$m (Figure~\ref{fig:h2}). The \arii\, line is visible in the original spectrum, while \neii\, sits right on top of the R-branch of C$_2$H$_2$ emission, which is clearly identified after subtracting the model spectra of hydrocarbons. This is the first time that \arii\, line is detected in a disk around a very low-mass star. The \neii\, line was only found in 4 late-type star disks with young ages of 1--2\,Myr \citep{Pascucci2013, Xie2023}, making this the first-ever detection for a much older disk ($\sim$34\,Myr old). 
Like H$_2$, both atomic lines are unresolved with MIRI. Emission from those ionized atoms could trace the disk surface gas that is heated by stellar high energy photons, and has been identified as disk wind/jet tracers (e.g., \citealt{vanBoekel2009, Pascucci2009, Bajaj2024, Barsony2024}), but no shift of line center is found in the spectra of J0446B with the limited spectral resolution of MIRI.

\begin{deluxetable*}{cccccc}[!th]
\tabletypesize{\scriptsize}
\tablecaption{Comparison of Properties among JWST/MIRI M-dwarf Sample\label{tab:mdwarf-comp}}
\tablewidth{0pt}
\tablehead{ 
\colhead{\bf Property}  & \colhead{\bf Sz\,114} & \colhead{\bf ChaI-147} & \colhead{\bf Sz\,28}& \colhead{\bf J160532} & \colhead{\bf J0446B} } 
\startdata
Region/Age (Myr) & Lupus/1--2 &  ChaI/2--3 & ChaI/2--3 & UpperSco/5--11 & $\chi^{1}$ For/34  \\
Spectral Type & M5  & M5.5 & M5.5 & M5 & M4.5 \\
$M_{*}\,(M_{\odot})$ & 0.16 & 0.11 & 0.12 & 0.16 & 0.18 \\
D (pc) & 157 &189 & 192 & 152 & 82 \\
$L_{*}\,(L_{\odot})$ & 0.2 & 0.03 & 0.03 & 0.04 & 0.016 \\ 
$\dot{M}\,(M_{\odot}$ yr$^{-1})$ & 7.9$\times10^{-10}$ & 7.0$\times10^{-12}$ & 1.8$\times10^{-11}$ & $10^{-10}\sim10^{-9}$ &  2.5$\times10^{-11}$ \\ 
$L_{X}$\,(erg \,s$^{-1})$ & 9.6$\times10^{29}$ & $<$1.1$\times10^{29}$ & 7.9$\times10^{27}$ & .. & 1.9$\times10^{28}$\\
$M_{\rm dust}\,(M_{\earth})$ & 30.1 & $<$0.22 & $<$0.35 & $<$0.18 & 0.1 \\
\hline
\hline
CH$_3$ (16.48) & $\times$ & $\checkmark$ & $\checkmark$ & $\times$ & $\checkmark$ \\ 
CH$_4$ (7.66) & $\times$ & $\checkmark$ & $\checkmark$ & $\sim\checkmark$ & $\checkmark$ \\ 
C$_{2}$H$_{2}$ (13.71) & $\checkmark$ & $\checkmark$ & $\checkmark$ & $\checkmark$ & $\checkmark$ \\  
$^{13}$CCH$_{2}$ (13.73) & $\times$ & $\checkmark$ & $\checkmark$ &$\checkmark$ & $\checkmark$ \\
C$_2$H$_4$ (10.53)& $\times$ & $\checkmark$ & $\times$ & $\times$ & $\checkmark$ \\  
C$_3$H$_4$ (15.80) & $\times$ & $\checkmark$ & $\checkmark$ & $\times$ & $\checkmark$ \\  
C$_2$H$_6$ (12.16) & $\times$ & $\checkmark$ & $\checkmark$ & $\times$ & $\checkmark$ \\  
C$_4$H$_2$ (15.92) & $\times$ & $\checkmark$ & $\checkmark$ & $\checkmark$ & $\checkmark$ \\
C$_6$H$_6$ (14.85) & $\times$ & $\checkmark$ & $\checkmark$ & $\checkmark$ & $\checkmark$ \\
HCN (13.99) & $\checkmark$ & $\checkmark$ & $\checkmark$ & $\checkmark$ & $\checkmark$ \\  
HC$_3$N (15.08) & $\times$ & $\checkmark$ & $\checkmark$ & $\times$ & $\checkmark$ \\ 
\hline
CO$_2$ (14.97) & $\checkmark$ & $\checkmark$ & $\checkmark$ & $\checkmark$ & $\checkmark$ \\ 
H$_2$O &  $\checkmark$ & $\times$ &  $\times$ &  $\sim\checkmark$ &  $\sim\checkmark$ \\
H$_2$ & $\checkmark$ & $\times$ & $\times$ & $\checkmark$ & $\checkmark$ \\
\neii\,(12.814) & $\checkmark$ & $\times$ & $\times$ & $\times$ & $\checkmark$ \\
\hline
refs & 1,2,3,4 & 2,5,6 & 5,7,8 & 9,10,11 & 12,13,14 \\ 
\enddata
\tablecomments{The energy ranges for quoted $L_{x}$: Sz\,114 (0.3--2\,keV); ChaI-147 (0.5--2\,keV); Sz\,28 (0.5--2\,keV); J0446B (0.1--1\,keV).}
\tablerefs{1: \citet{Ansdell2016}; 2: \citet{Manara2023}; 3: \citet{Gondoin2006}; 4: \citet{Xie2023}; 5: \citet{Pascucci2016}; 6: \citet{Arabhavi2024Sci...384.1086A}; 7: \citet{Kanwar2024arXiv}; 8: \citet{Feigelson2004}; 9: \citealt{Barenfeld2016}; 10: \citet{Pascucci2013}; 11: \citet{Tabone2023NatAs...7..805T}; 12: \citet{Silverberg2020}; 13: \citet{Laos2022}; 14: this work}
\end{deluxetable*}

\section{Discussion} \label{sec:diss}

\subsection{J0446B as a long-lived primordial disk}
Recently, about ten disks in systems older than 20 Myr have been identified with accretion signatures (e.g., \citealt{Boucher2016, Murphy2018, Silverberg2020, Gaidos2022}), indicating the presence of gas at later stages than previously found. Notably, all these disks are associated with low-mass stars and brown dwarfs (Spectral Type later than M4).
One straightforward explanation for these systems is that low-mass stars may dissipate their primordial disks more slowly than their higher-mass counterparts. Infrared photometric measurements suggest that dust disks around lower mass stars could indeed have longer lifetimes than those around more massive stars (e.g., \citealt{Carpenter2006, Ribas2015}). However, a surviving period of 30--50\,Myr (5--10 times longer than the typical disk lifetime) was largely unexpected. Alternative explanations for these old accreting disks include giant collisions between planetesimals \citep{Flaherty2019}, dynamical interactions between stars and their disks, and tidal disruptions caused by giant planets \citep{Silverberg2020}, while it remains unclear why these scenarios would be more common for low-mass stars.

The detection of spatially unresolved H$_2$ and \neii\,lines strongly suggests that J0446B hosts a long-lived primordial gas disk, marking the first confirmed case of disk gas surviving for more than 30\,Myr. A survey of debris disks using Spitzer set stringent upper limits on the H$_2$ column density in the region where terrestrial planets form, showing it to be less than 1/10,000th of the Minimum Mass Solar Nebula \citep{Pascucci2006}.
Recent JWST observations towards the well-studied debris disk around the 23-Myr-old $\beta$ Pic identified \arii\,line emission, which is about 10$\times$ brighter than our detection in J0446B, but did not detect H$_2$ and \neii\,lines at comparable levels \citep{Worthen2024}.
This situation parallels the mechanism explaining the noble gas content in Jupiter's atmosphere: Ar can be trapped in water ice within the cold outer protoplanetary disk and later be released into the gas phase \citep{Monga2015, Wu2024arXiv}. 
The presence of highly volatile species of H$_2$ and Ne in J0446B thus indicates it retains primordial gas. 
In the case of J0446B, the comparable fluxes of the detected \neii\,and \arii\,lines suggest that soft X-ray/EUV radiation is the main ionizing source \citep{Szulagyi2012}. This indicates that high-energy photons from the central low-mass star may contribute to the heating and ionization of the disk, but they may not be sufficient to completely disperse the disk. 

Table~\ref{tab:mdwarf-comp} summarizes the line detections for the five mid-to-late M-star disks recently analyzed with JWST MIRI/MRS spectra. The inner disk of the $\sim$34-Myr-old star J0446B contains all the molecular species detected in younger systems of 3--10\,Myr old. Furthermore, the excitation conditions for hydrocarbons ($T\sim250-300$\,K and $R_{\rm slab}\sim0.05-0.1$\,AU) align closely with those observed in ChaI-147 \citep{Arabhavi2024Sci...384.1086A} and Sz\,28 \citep{Kanwar2024arXiv}, two similarly hydrocarbon-rich systems. These results further support the idea that J0446B hosts a primordial gas disk.

\subsection{High $L_{\rm C{_2}H{_2}}$/$L_{\rm H{_2}O}$ ratio in disks around low-mass stars}
As revealed by our MIRI/MRS observations, the inner gas disk of the M4.5 star J0446B is dominated by molecular emission from hydrocarbons (see Section~\ref{sec:results}). Earlier studies with Spitzer/IRS faced limitations in accessing disks around late-M stars due to their faintness. However, the stronger C$_2$H$_2$ over HCN line fluxes in late-M star disks \citep{Pascucci2009}, along with the weak/absent H${_2}$O emission \citep{Pascucci2013}, have suggested enhanced carbon chemistry, as compared to their solar analogs. 
Recent JWST results have corroborated this finding, revealing detections of large numbers of hydrocarbon molecules in low-mass star disks (e.g., \citealt{Tabone2023NatAs...7..805T, Arabhavi2024Sci...384.1086A, Kanwar2024arXiv}), with the exception of Sz 114 \citep{Xie2023}.

\begin{figure*}[!t]
\centering
    \includegraphics[width=0.48\textwidth]{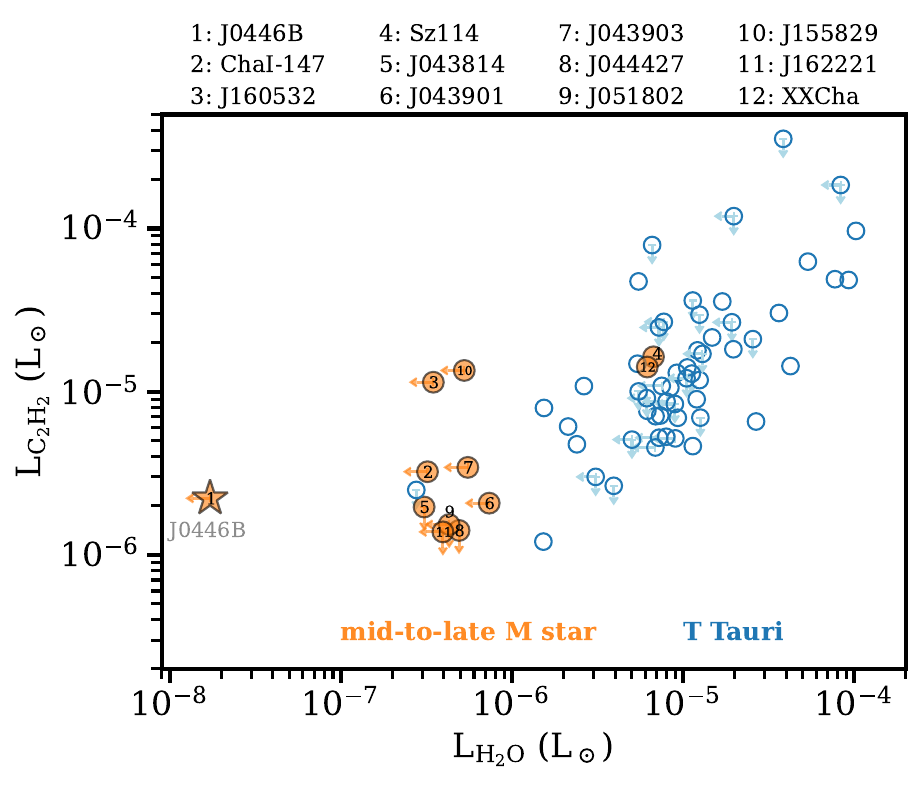}
    \includegraphics[width=0.48\textwidth]{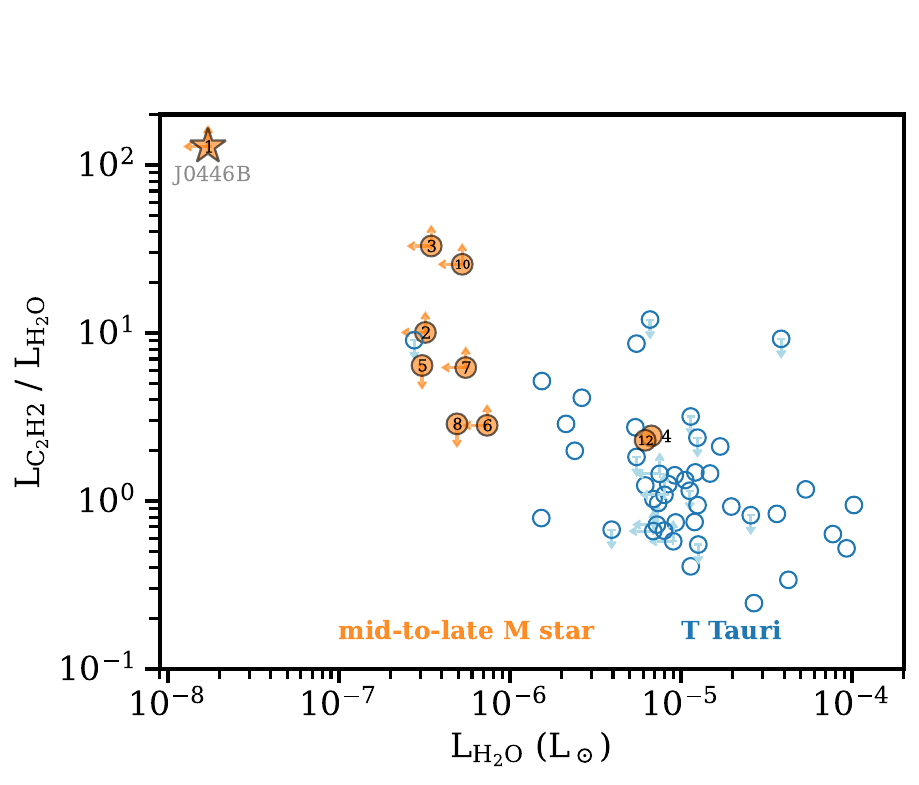}
\caption{{\bf\textit{Left:}} the 17.22\,$\mu$m water line luminosity vs. C$_2$H$_2$ $Q$-branch line luminosity for a mixed sample from JWST and Spitzer; {\bf\textit{Right:}} the 17.22\,$\mu$m water line luminosity vs. C$_2$H$_2$ to water flux ratio. The M star sample refers stars with spectral type later than M3, and is labelled with the target name: 1 - this work, 2 - \citet{Arabhavi2024Sci...384.1086A}, 3 - \citet{Tabone2023NatAs...7..805T}, 4 - \citet{Xie2023}, 5 to 11 - \citet{Pascucci2013}, 12 and all T Tauri disks from \citet{Banzatti2020}. The 3$\sigma$ upper limits are indicated with arrows. Targets with upper limits for both lines are excluded in the right panel for flux ratios. \label{fig:linefluxcomp}}
\end{figure*}

Figure~\ref{fig:linefluxcomp} demonstrates the distribution of mid-infrared line luminosities ($4 \pi d^2 F$) and their ratios for disks around different types of host stars. 
The comparison here focuses on C$_2$H$_2$ and H$_2$O, as both were often detected in previous Spitzer/IRS observations and serve as key carriers for carbon and oxygen in disk gas, respectively. The T Tauri disk sample is taken from \citet{Banzatti2020}, which includes tabulated line fluxes. 
The mid-to-late M disk sample (with SpTy later than M3) comprises J0446B from this work, three disks with newly published JWST spectra (Sz\,114, ChaI-147, J160532), seven disks from \citet{Pascucci2013}, and XX\,Cha from \citet{Banzatti2020}. 
For disks with no reported line fluxes, we calculated the C$_2$H$_2$ line flux using published slab model spectra and integrated over a similar wavelength range (13.65--13.72\,$\mu$m) as \citet{Banzatti2020}. For water, we followed \citet{Xie2023} that only considered the 17.22$\mu$m feature and adopted their updated water fluxes for the sample in \citet{Banzatti2020}. 

M-star disks have generally fainter mid-infrared lines than those in T Tauri disks, which aligns with the expectation that line luminosity scales with disk heating (from both stellar and accretion luminosities, \citealt{Salyk2011, Banzatti2020}). The median value of $L_{\rm H{_2}O}$ in the two samples differs by a factor of 15, while this difference for $L_{\rm C{_2}H{_2}}$ is only $\sim$4. Consequently, M-star disks tend to have higher $L_{\rm C{_2}H{_2}}$/$L_{\rm H{_2}O}$, ranging from 2 to $\sim$100, while this ratio is typically around 1 for T Tauri disks. We note that, for many M-star disks, this ratio may be even underestimated due to the non-detection of water. 
Our target disk J0446B (highlighted in Figure~\ref{fig:linefluxcomp}) shows the largest line ratio of $L_{\rm C{_2}H{_2}}$/$L_{\rm H{_2}O}$.

\begin{figure*}[!th]
\centering
    \includegraphics[width=0.98\textwidth]{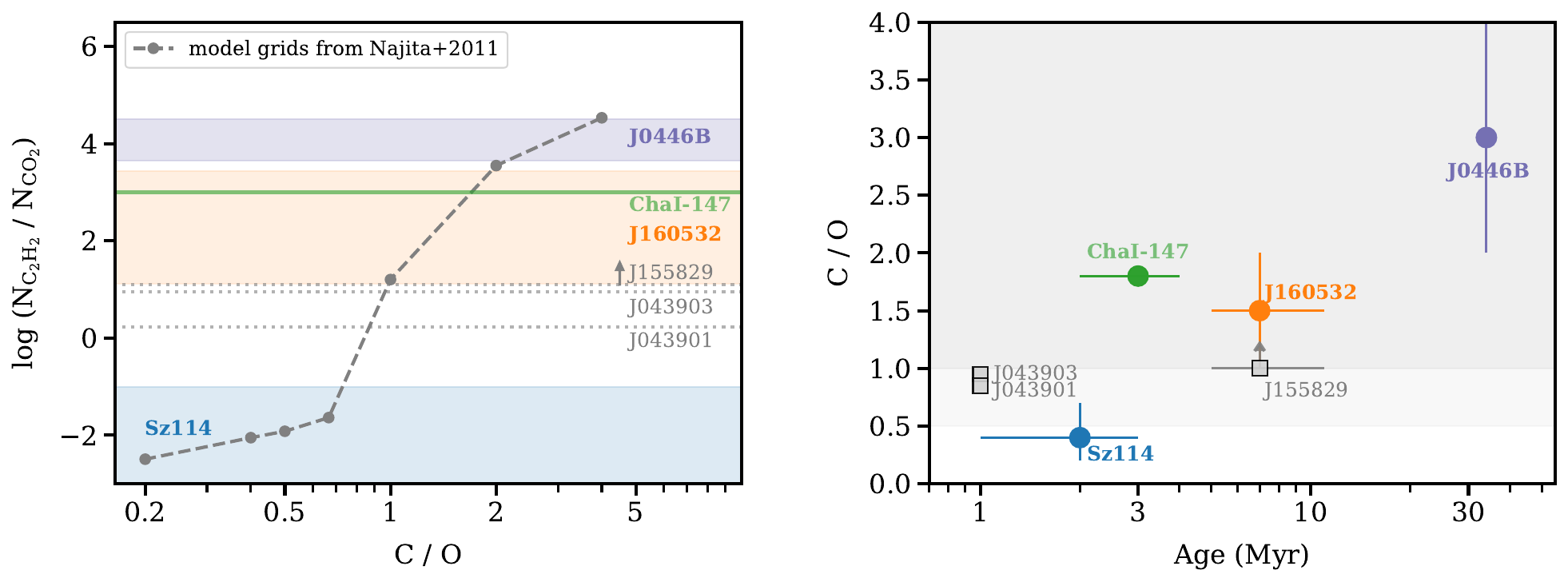}
\caption{{\bf\textit{Left:}} The input gas C/O ratio vs. the output column density ratio of C$_2$H$_2$ and CO$_2$ from \citet{Najita2011_model} chemical models (connected grey dots). The observed values/ranges of the JWST mid-to-late M-star disk sample are highlighted in colors, except for Sz\,28, for which the derived column densities are highly uncertain \citep{Kanwar2024arXiv}. Three disks with derived column densities from Spitzer/IRS data \citep{Pascucci2013} are indicated as grey dotted lines.  {\bf\textit{Right:}} Stellar age vs. the derived C/O ratio from model grids. Lower limit is marked for J155829 as CO$_2$ was not detected with the Spitzer data. \label{fig:c2oratio}}
\end{figure*}

\subsection{The evolution of inner disk C/O ratio}
Within the mid-to-late M-star disk sample, the disks around Sz\,114 and XX\,Cha are two clear outliers, whose line properties are more aligned with the T Tauri sample (marked as number 4 and 12 in Figure~\ref{fig:linefluxcomp}). The improved sensitivity of JWST now allows us to detect fainter lines. As summarized in Table~\ref{tab:mdwarf-comp}, at least 14 molecular species are currently identified in M-star disks. However, only 3 species (C$_2$H$_2$, HCN, and CO$_2$) are commonly detected among the five low-mass star disks with recently published JWST MIRI/MRS spectra. 
ChaI-147 \citep{Arabhavi2024Sci...384.1086A}, Sz\,28 \citep{Kanwar2024arXiv}, J160532 \citep{Tabone2023NatAs...7..805T}, and J0446B (this work) all contain a large number of hydrocarbon molecules but show no/weak H$_2$O emission. In contrast, the only other molecular detection in Sz\,114 is H$_2$O (excluding H$_2$, \citealt{Xie2023}; JWST data for XX\,Cha has not been published yet).


\subsubsection{The estimate of inner disk C/O ratio}
Thermo-chemical models that account for high gas-phase C/O ratios often predict high columns of hydrocarbons (e.g., \citealt{Najita2011_model, Woitke2018, Anderson2021}). Specifically, \citet{Najita2011_model} investigated how molecular column densities vary on the inner disk surface across a wide range of C/O ratios, from 0.2--4, with solar value taken as 0.45. 
Those models show that increasing the C/O ratio to above 0.7 can lead to a sharp rise (or drop) in the column density of C$_2$H$_2$ (or H$_2$O), and this difference can be three orders of magnitude when reaching a C/O ratio of $\sim$\,1. Therefore, the column density ratio of carbon- and oxygen-bearing molecules can be expected to reflect the C/O ratio, though other physical parameters (e.g., dust-to-gas ratio, inner disk structure) may also regulate the observed emission. 

For the M-star disk sample analyzed with JWST spectra, where CO$_2$ is the robustly detected oxygen-bearing molecule, our estimate of the C/O ratio is based on the slab model results of C$_2$H$_2$ and CO$_2$. 
Considering the total column of C$_2$H$_2$ (predominately the optically thick component), the column density ratios in the hydrocarbon-rich disks range from $\sim10$ to 10000, yielding high C/O ratios of 1 to 4 (Figure~\ref{fig:c2oratio}). In contrast, the water-rich disk around Sz\,114 exhibits a significantly lower column of C$_2$H$_2$ compared to CO$_2$, with the disk gas C/O ratio approximately at the solar value. Even when considering the optically thin component of C$_2$H$_2$ in these hydrocarbon-rich disks, the column density ratio remains generally higher than that observed in Sz\,114. 
Figure~\ref{fig:c2oratio} also includes the three late-M star disks with retrieved column densities based on Spitzer spectra by \citet{Pascucci2013}, which resulted in relatively high C/O ratios of $\sim$1 (lower limit for the Upper Sco target J155829 with CO$_2$ non-detection). Recently, \citet{Grant2023} found significant inconsistencies in the inferred CO$_2$ column density of a T Tauri disk as derived from Spitzer and JWST spectra. Thus, future high-quality JWST data would be needed to provide more reliable estimates. When considering the marginal detection of water in J0446B, the resulting C/O ratio is also close to 2 based on the column density ratio of C$_2$H$_2$ and H$_2$O.

We note several caveats regarding the above estimates of the inner disk gas C/O ratio. The chemical models presented in \citet{Najita2011_model} were developed for T Tauri disks, which are overall warmer than disks around lower mass stars studied here. The higher temperature of model disks may suppress CO$_2$ formation \citep{Bosman2022}, potentially under-predicting the CO$_2$ column density in M-star disks. Moreover, their chemical network did not account for pathways to more complex hydrocarbons (beyond those with up to two carbon atoms), which likely resulted in an overestimation of C$_2$H$_2$. Together, for a given C/O ratio, we would expect a lower column density ratio of C$_2$H$_2$ and CO$_2$ in disks around the lower-mass stars than that tabulated in \citet{Najita2011_model}. These models therefore likely underestimated the C/O ratio for the M disk sample. 
A recent modeling work by \citet{Kanwar2024arXiv} investigated the inner disk chemistry around very low-mass stars with two different C/O ratios of 0.45 and 2, utilizing a more complete chemical network. They also identified five orders of magnitude differences in the column density ratio of C$_2$H$_2$ and CO$_2$ between the two scenarios. This variation arises from the efficient formation of hydrocarbons through neutral-neutral and ion-molecule pathways when there is excess carbon. 
Future works that explore a wider parameter space for disks around very low-mass stars are certainly needed to better interpret observations. For now, the models from \citet{Najita2011_model} provide a first-order comparison.

\begin{figure*}[!th]
\centering
    \includegraphics[width=0.95\textwidth]{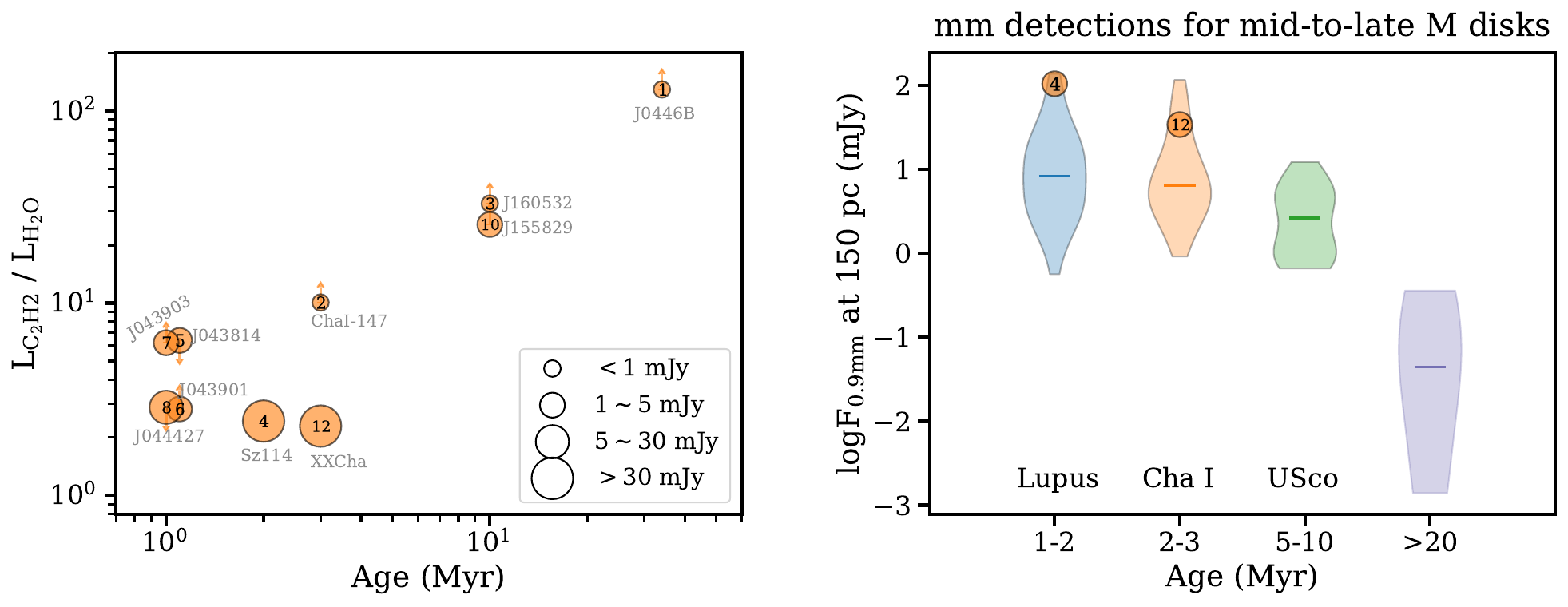}
\caption{{\bf\textit{Left:}} stellar age vs. C$_2$H$_2$ to water flux ratio for the mid-to-late M disk sample, using the same target labelling as Figure~\ref{fig:linefluxcomp}. The size of the symbol corresponds to the disk flux at 0.9\,mm (scaled to 150\,pc). {\bf\textit{Right:}} the distribution of mm fluxes for mid-to-late M-star disks from Lupus \citep{Ansdell2016}, Cha I \citep{Pascucci2016}, Upper Sco \citep{Barenfeld2016}, and other older systems (\citealt{Flaherty2019, Cronin-Coltsmann2023}, this work). Values for young disks are taken from the compiled table in \citet{Manara2023}, and only $>3\sigma$ detections are included. The width in each ``violin" reflects the density of data points along the y-axis. \label{fig:mm-agecomp}}
\end{figure*}

Figure~\ref{fig:c2oratio} (right panel) shows the distribution of the estimated C/O ratio as a function of stellar age. Considering the large uncertainties in age estimates of young stars, here we use the average values of the corresponding star clusters.
The current JWST sample of disks around very low-mass stars covers a broad age range from $\sim$\,2 to 35\,Myr, albeit with a very small sample. Within this sample, the O-rich disk around Sz\,114 from Lupus stands out as the youngest, which is further supported by its high stellar luminosity (see Table~\ref{tab:mdwarf-comp}). The two young Taurus disks with measurements from Spitzer (J043901, J043903, \citealt{Pascucci2013}) are more C-rich than Sz\,114 but not as extreme as the older systems. As shown in Figure~\ref{fig:mm-agecomp} (left panel), the same trend holds when including the remaining mid-to-late M-star disks that lacked column density estimates and using the line ratio of $L_{\rm C{_2}H{_2}}$/$L_{\rm H{_2}O}$ as an approximation. This comparison suggests that young disks exhibit a range of C/O ratios, while older systems appear to be mostly very C-rich.

\subsubsection{The mechanisms of disk chemical evolution}
What are the physical processes that drive the inner disk's chemical evolution? The destruction of carbonaceous grains and PAHs has been proposed as one potential approach to enhance the carbon content in the inner disk \citep{Tabone2023NatAs...7..805T, Colmenares2024arXiv}. Depending on the composition and internal structure of carbonaceous materials, their thermal sublimation can occur at temperatures of either 500\,K or 1200\,K \citep{Gail2017, Li2021Sci}. However, for three out of the four C-rich disks analyzed with JWST spectra, the excitation temperatures for hydrocarbons are consistently low, clustered around $\sim$300\,K. This suggests that carbon grain sublimation may not be the primary mechanism responsible for the observed excess carbon in these disks. 
In addition, the photo-destruction of carbon grains is unlikely to be significant in disks around late-type stars with weak radiation fields. The general absence of PAH emission features is consistent with their weak UV field. \citet{Tabone2023NatAs...7..805T} also suggested that grain growth and settling, as indicated by the lack of 10\,$\mu$m silicate dust feature in the disk around J160532, may have increased the UV penetration depth and facilitated grain destruction. However, other disks exhibit a range of dust emission levels (clear silicate emission in the J0446B disk and even stronger feature in Sz\,28).

Alternatively, as suggested by \citet{Mah2023}, it is the combination of two processes happening faster in disks around lower-mass stars that affect the inner disk composition: 1) the inward drift and subsequent sublimation of O-rich icy pebbles; 2) the accretion of outer disk C-rich gas to the inner region.
Icy pebbles in the outer region of a smooth disk are subject to inward radial drift due to the aerodynamic gas drag, and the associated timescale is shorter for disks around lower mass stars \citep{Pinilla2013}. 
Due to the relatively close-in ice lines of H$_2$O and CO$_2$ (major oxygen-bearing ices), the sublimation of pebbles coated with those ices will initially increase the oxygen abundances in the inner disk until the O-rich gas is accreted onto the star. This is supported by the recent finding of excess cold water emission in small disks that likely experienced more efficient pebble drift \citep{Banzatti2023}. On the other hand, the initial gas phase C/O ratio is expected to be higher at larger disk radii, primarily due to the differing snowlines of carbon- and oxygen-bearing ices \citep{Oberg2011}. Bright C$_2$H emissions observed at tens of au disk from the host stars, as detected by ALMA, also suggest a high C/O ratio for the outer disk gas (for both T Tauri and mid-to-late M stars, e.g., \citealt{Bergin2016, Pegues2021}). When accretion drives the outer C-rich gas inward, while water vapour from the inward drifting icy pebbles gets accreted onto the star, the inner disk C/O ratio gradually increases. The models of \citet{Mah2023} showed that pebble drift occurs faster than gas accretion and the inner disk around a star with 0.1--0.3\,$M_{\odot}$ can reach the super-solar C/O ratio within 1\,Myr for a smooth disk, assuming $\alpha$ viscosity is 10$^{-3}$. However, given the same viscosity, this transition only occurs around 3\,Myr for disk around a 1\,$M_{\odot}$ star because of the further out ice line locations.

Due to the fast grain evolution in disks around very low-mass stars, pebbles in these disks are likely to have been largely depleted due to inward migration by the typical protoplanetary disk ages ($>$1\,Myr), and the O-rich stage may have already passed. 
This is consistent with the very faint to non-detection of mm emission in the four C-rich disks around very low-mass stars ($\lesssim$1\,mJy, see Table~\ref{tab:mdwarf-comp}). The water-dominated disk around Sz\,114 is, however, very bright in the mm wavelength ($>$100\,mJy at 0.9\,mm, scaled to a common distance of 150\,pc, \citealt{Ansdell2016}), with a shallow dust gap visible at $\sim$39\,au \citep{Huang2018} and another feature around 10\,au inferred from super-resolution fitting \citep{Jennings2022}. The presence of these dust substructures could have reduced the efficiency of pebble drift \citep{Kalyaan2023}, thereby prolonging the pebble disk lifetime and the oxygen supply to the inner disk \citep{Xie2023}. This picture also holds when extending to the Spitzer and JWST combined sample. As shown in Figure~\ref{fig:mm-agecomp} (left panel), the two disks with the brightest mm emission (Sz\,114 and XX\,Cha) exhibit the lowest $L_{\rm C{_2}H{_2}}$/$L_{\rm H{_2}O}$ ratios (a proxy for C/O ratio). The third brightest disk (number 8: J044427, \citealt{Ricci2014}) also has a comparably low line ratio (upper limit). On the other hand, disks with high line ratios (e.g., $>5$) all have faint mm emission ($F_{\rm mm}<5$\,mJy). We note the fast depletion of dust pebbles is not in conflict with the high frequency of terrestrial planets detected around low-mass stars, if those planets form very early on and are not capable of blocking pebble drift.

The violin plot in the right panel of Figure~\ref{fig:mm-agecomp} illustrates the distribution of mm emission for mid-to-late M-star disks across different system ages. Both the maximum and median $F_{\rm mm}$ decrease with age. Based on the pebble drift scenario, maintaining an inner O-rich status requires a continuous influx of icy pebbles from the outer disk region, thus this characteristic is more likely to be observed in younger systems. This is consistent with the fact that the two water-rich disks, Sz\,114 and XX Cha, are from younger regions. In particular, among the late-type star disks, Sz\,114 ranks as the second brightest in Lupus \citep{Ansdell2016}, and XX Cha is among the top 15\% of Cha I \citep{Pascucci2016}. Given the generally faint mm emission in disks around very low-mass stars, O-rich disks are expected to be exceptions rather than the norm, with their fraction likely decreasing over time. Additionally, mm-bright disks are not necessarily O-rich if a strong dust trap, which completely blocks the inward pebble migration, forms early in the disk \citep{Mah2024}. This further reduces the occurrence of O-rich disks, making carbon chemistry more characteristic of disks around very low-mass objects, particularly in later evolutionary stages.

To explore the evolution of the C/O ratio in the disk of J0446B, we followed an approach similar to \citet{Mah2023} for a smooth disk but extended the time evolution to 40\,Myr, whereas the endpoint was set to 10\,Myr in \citet{Mah2023}. We used the code \texttt{chemcomp} \citep{Schneider2021} that models the transport,  sublimation, and condensation of solid and gas species in a 1D viscous $\alpha$-disk \citep{Shakura1973, Lynden-Bell1974}. The evolution of dust grain was computed with the two-population algorithm of \citet{Birnstiel2012}. Thus, \texttt{chemcomp} can simulate the evolution of the solid and gas-phase composition of the disk, considering both pebble drift and viscous evolution. The stellar mass ($0.1\mathrm{~M_\odot}$) and stellar luminosity ($0.02\mathrm{~L_\odot}$) were set to match J0446B. The initial characteristic gas disk radius was set to $R_\mathrm{c} = 55 \mathrm{~AU}$, consistent with measured disk sizes around very low-mass stars \citep{Kurtovic2021, Shi2024}. The dust fragmentation velocity was fixed at $v_\mathrm{frag} = 5 \mathrm{~m~s^{-1}}$. Similarly to \citet{Mah2023}, we allocated $60\%$ of the carbon content in refractory grains, $20\%$ in CO, $10\%$ in CO$_2$, and $10\%$ in CH$_4$. The output gas-phase C/O ratio, considering the gas surface density of all carbon and oxygen-bearing molecules, was calculated every $0.05 \mathrm{~Myr}$. 

\begin{figure}[!th]
\centering
    \includegraphics[width=0.98\linewidth]{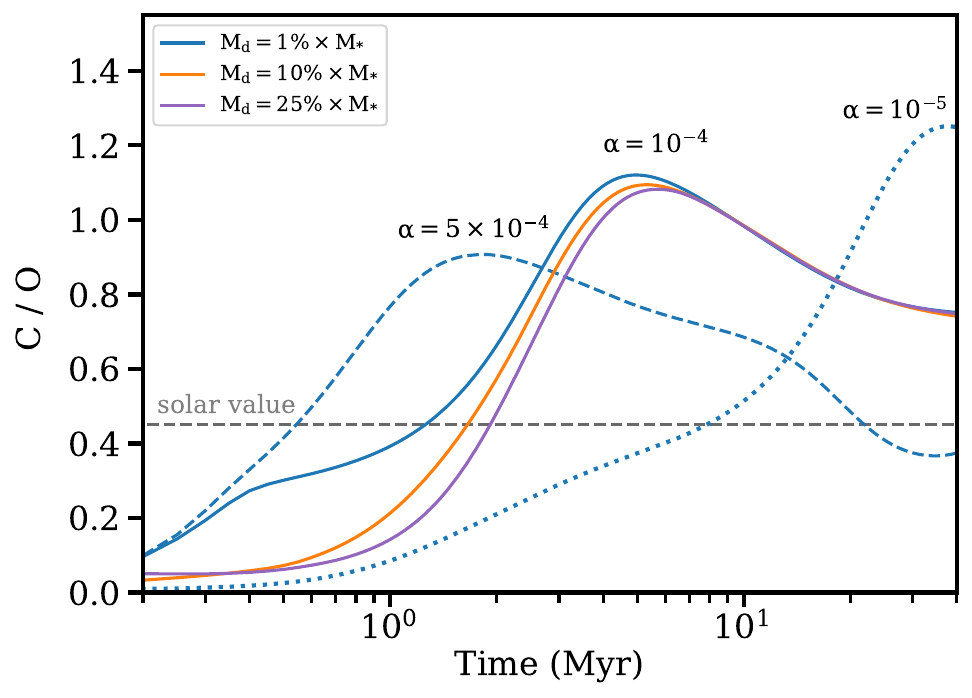}
\caption{The modelled time evolution of gas-phase C/O ratio using \texttt{chemcomp} \citep{Schneider2021} in a disk around a very low-mass star, with stellar properties consistent with J0446B. The line styles represent different $\alpha$ viscosity and the line colors represent different initial disk masses. \label{fig:model_c2o}}
\end{figure}

Figure~\ref{fig:model_c2o} compares the time evolution of gas-phase C/O ratio with varying disk masses and $\alpha$ viscosity. The super-solar C/O status can emerge very early and persist for an extended period until the disk gas largely dissipates. The onset and duration of this phase are highly sensitive to disk viscosity, which strongly influences the efficiency of grain growth and the radial transport of both icy pebbles and gas \citep{Birnstiel2012, Pinilla2021}. Thus, maintaining a high C/O ratio at $\sim$30\,Myr requires a rather low $\alpha-$viscosity ($< 10^{-4}$). This is consistent with current theoretical understanding of how to sustain a primordial disk over tens of Myr, considering also low external photoevaporation \citep{Coleman2020, Wilhelm2022}.

\subsection{The implications for planet formation}
Extended disk lifetimes have significant implications for planet formation and evolution. Firstly, a long-lived disk can provide a prolonged period for building planetary cores under the core-accretion paradigm, which may alleviate the current tension of the need for early planet formation (e.g., \citealt{Mulders2021}). 
Secondly, the prolonged presence of gas in the disk can impact the dynamical evolution of forming planets. Disk gas damps their orbital eccentricities and facilitates planet migration, which results in compact planetary systems with resonant chains (e.g., \citealt{Zawadzki2021}). This dynamic process may help explain the architecture of the TRAPPIST-1 system, which hosts seven Earth-like planets within 1 au around a late M star, arranged in near-resonant orbits \citep{Gillon2017}.

From the chemical perspective, the primordial atmospheres of planets forming around very low-mass stars will likely inherit super-Solar C/O ratios. An elevated C/O ratio could alter the atmospheric mean molecular weight and thermal structure (e.g., \citealt{Madhusudhan2012, Moses2013}), potentially impacting subsequent atmospheric loss. In addition, a primordial atmosphere abundant in organic compounds could generate organic hazes \citep{Bergin2023}, which may have a lasting impact on the detectability of spectral features and its chemical evolution, as suggested e.g., by Titan in our Solar System (e.g., \citealt{Horst2017}). 
Furthermore, if a large fraction of carbon is in the gas phase and gradually lost over time, terrestrial planets assembled from dust solids with complementary composition could instead be carbon-poor, even under disk conditions where significant refractory carbon should be retained. This scenario forms an alternative to models in which early assembled, volatile poor building blocks dominate terrestrial planet assembly \citep{Li2021Sci}, if our carbon-depleted Earth was formed later within the solar nebula. 
The trends established through JWST/MIRI spectroscopy (by extending to Sun-like stars) can also help place the volatile chemistry of carbonaceous chondrites, which formed with 2.5--3.5\,Myr after the early protosolar phase, into broader context. 
The structural nature of Insoluble Organic Matter (IOM), the primary carbon reservoir in carbonaceous chondrites, is dominated by a mixture of aliphatic groups mixed with small aromatic species \citep{Derenne2010}.  
This, and the presence of stable diradical species suggests that IOM was assembled from precursors in warm/hot regions of the disk. Both the nature of the hydrocarbon-rich spectra observed here, and the expected temporal evolution of the inner disk C/O ratio, are consistent with this interpretation, and suggest the features we observe here are broadly relevant to planetary system formation.

\section{Summary} \label{sec:summ}
In this paper, we presented the JWST MIRI/MRS spectrum of the inner disk of J0446B, an M4.5 star that is $\sim$34\,Myr old but shows hints of accretion. The high-quality JWST spectrum unveiled a wealth of molecular lines within this disk, allowing for the first detailed characterization of gas-rich disks at such an advanced age. Slab models were used to constrain the inner disk gas properties, which were then compared with those of young disks around stars with similar and earlier spectral types to explore the disk chemical evolution. Our main findings are summarized as follows: 

\begin{enumerate}

\item The spectrum is dominated by emission lines from hydrocarbon molecules, including the detection of 9 such species: CH$_3$, CH$_4$, C$_2$H$_2$, $^{13}$CCH$_2$, C$_2$H$_4$, C$_2$H$_6$, C$_3$H$_4$, C$_4$H$_2$, and C$_6$H$_6$. The two bumps around $\sim$7 and 14\,$\mu$m correspond to a highly optically thick ($N\sim10^{22}$ cm$^{-2}$) component of C$_2$H$_2$. Other hydrocarbons share similar excitation conditions, with $T\sim250-300$\,K and $R_{\rm slab}\sim0.05-0.1$\,AU.

\item Additional detected lines include two N-bearing molecules of HCN and HC$_3$N, two O-bearing molecules CO$_2$ and $^{13}$CO$_2$, five H$_2$ pure rotational transitions, and two atomic lines of \neii\, and \arii. All of these lines represent the first detections in disks at the ages of $\sim$30\,Myr. Line emission from H$_2$O is only marginally detected, with a total emitting water molecule number of $3.5\times10^{42}$ (corresponding to $1.8\times10^{-8}\,M_{\earth}$).

\item The \arii\, line is detected for the first time in disks around very low-mass stars. The flux ratio of \neii\,and \arii\,lines suggest that soft X-ray/EUV radiation might be the main ionizing source in this disk.

\item The gas properties closely resemble those of young disks around stars with similar spectral types. Based on model grids of \citet{Najita2011_model}, we estimated a very high C/O ratio of $\gtrsim$2 in the disk of J0446B. Considering the combined effects of pebble drift and gas accretion, such a high ratio is expected at its old age if the $\alpha-$viscosity is very low ($\lesssim10^{-4}$).  
\end{enumerate}

These detections, especially the spatially unresolved lines of H$_2$ and Ne, strongly support the idea that primordial disks around very low-mass stars can persist for tens of Myrs. The long presence of gas in disks has wide implications for the final architectures of planetary systems. Our observations motivate future studies to quantify the frequency of long-lived disks and to further test the disk chemical evolution pathways.

\paragraph{Acknowledgments}
We thank the referee for careful reading of the manuscript and helpful suggestions. F.L. thanks Luca Matra for helpful discussions. Support for F.L. was provided by NASA through the NASA Hubble Fellowship grant \#HST-HF2-51512.001-A awarded by the Space Telescope Science Institute, which is operated by the Association of Universities for Research in Astronomy, Incorporated, under NASA contract NAS5-26555. 
A.B, K.P., C.X., and I.P. are partially supported by STScI Grant \#JWST-GO-03153.001-A. 
A portion of this research was carried out at the Jet Propulsion Laboratory, California Institute of Technology, under a contract with the National Aeronautics and Space Administration (80NM0018D0004). 
D.H. is supported by a Center for Informatics and Computation in Astronomy (CICA) grant and grant number 110J0353I9 from the Ministry of Education of Taiwan. D.H. also acknowledges support from the National Science and Technology Council, Taiwan (Grant NSTC111-2112-M-007-014-MY3, NSTC113-2639-M-A49-002-ASP, and NSTC113-2112-M-007-027). 
P.P. acknowledges funding from the UK Research and Innovation (UKRI) under the UK government’s Horizon Europe funding guarantee from ERC (under grant agreement No 101076489). 
G.J.H. is supported by grant IS23020 from the Beijing Natural Science Foundation.

This work includes observations made with the NASA/ESA/CSA James Webb Space Telescope. The JWST data used in this paper can be found in MAST: \dataset[10.17909/5d5a-es35]{http://dx.doi.org/10.17909/5d5a-es35}. 
This paper makes use of the following ALMA data: ADS/JAO.ALMA\#2021.1.00871.S. ALMA is a partnership of ESO (representing its member states), NSF (USA) and NINS (Japan), together with NRC (Canada), MOST and ASIAA (Taiwan), and KASI (Republic of Korea), in cooperation with the Republic of Chile. The Joint ALMA Observatory is operated by ESO, AUI/NRAO and NAOJ. The National Radio Astronomy Observatory is a facility of the National Science Foundation operated under cooperative agreement by Associated Universities, Inc.


\vspace{5mm}
\facilities{JWST(MIRI), ALMA}

\software{
         chemcomp \citep{Schneider2021},
          iris \citep{Munoz-Romero2023zndo}, 
          dynesty \citep{Speagle2020}
          }

\appendix

\section{J0446A} \label{sec:J0446A}
\begin{figure*}[!t]
\centering
    \includegraphics[width=0.9\textwidth]{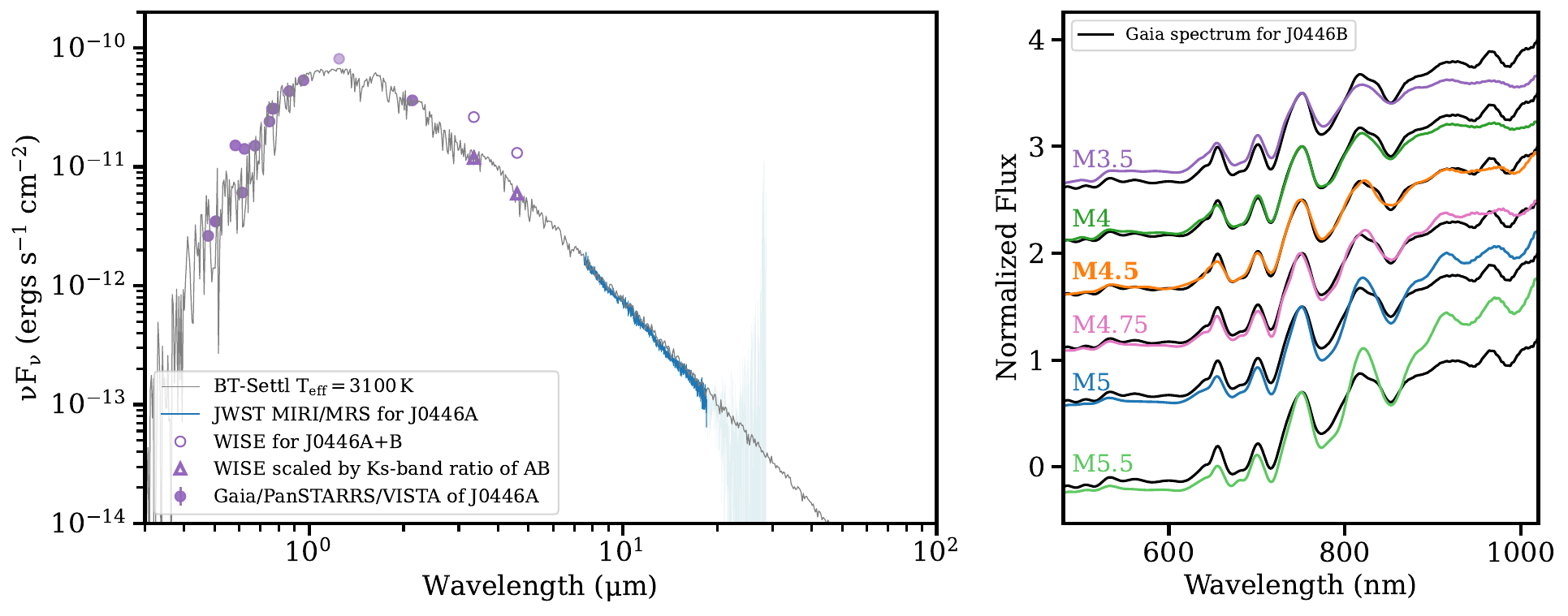}
\caption{\textbf{Left:} The spectral energy distribution of J0446A including the JWST MIRI/MRS spectrum in blue (the noisy long-wavelength range is marked with lighter color). The grey curve shows a stellar photospheric model with $T_{\rm eff}=3100\,K$. The VISTA J-band flux of J0446A might be problematic as this flux is comparable to that of J0446B, while J0446A is consistently fainter in all other bands. \textbf{Right:} The comparison of the Gaia XP spectra of J0446B to XP spectra of objects in TW Hya Association with varying spectral types. The molecular absorption bands are consistent with a spectral type of M4.5. \label{fig:sed_A}}
\end{figure*}

With a projected separation of 2\farcs3 from J0446B, the stellar companion J0446A lies partially outside the MIRI/MRS Channel 1 detector. Our extracted MIRI/MRS spectrum, beginning around 7.5\,$\mu$m, aligns well with a stellar photospheric model at $T_{\rm eff}=3100\,K$ \citep[][see Figure~\ref{fig:sed_A},]{Allard2012}, supporting our reassessment of the stellar properties discussed in Section~\ref{sec:target}. The ALMA non-detection of mm emission (3$\sigma$ upper limit of 0.08\,mJy) for J0446A is consistent with its photospheric-like spectrum. 
Although J0446A and J0446B are coeval and share similar stellar types, J0446A appears to have followed a different evolutionary path from J0446B. As previous studies have shown \citep[e.g.,][]{Barsony2024}, twin disks can evolve differently, though the reasons for this remain an open question.

\section{Stellar absorption features} \label{stellar_absorp}
\begin{figure*}[!t]
\centering
    \includegraphics[width=0.9\textwidth]{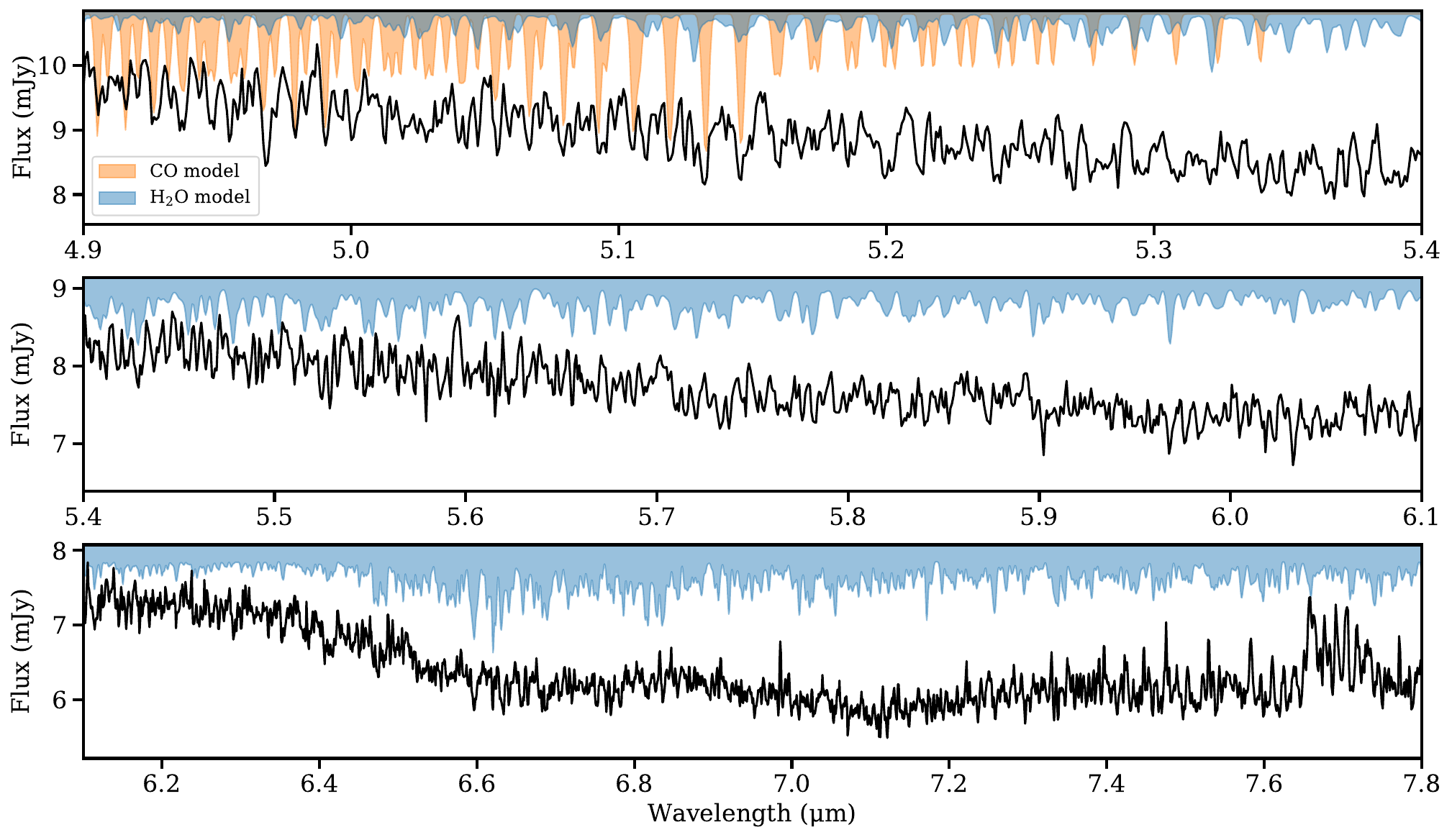}
\caption{The observed spectrum (black) in comparison with CO (orange) and H$_2$O (blue) models (scaled) with emitting temperature of 3100\,K. \label{fig:stellar}}
\end{figure*}

The short wavelength of the MIRI/MRS spectrum of J0446B is strongly affected by contributions from the central star. 
Figure~\ref{fig:stellar} compares slab models of CO and H$_2$O at a temperature of 3100\,K, consistent with the stellar effective temperature, to the observed spectrum. Absorption features from CO dominate from 4.9--5.3\,$\mu$m, and H$_2$O absorption can influence a much wider wavelength range. The dim around 6.6\,$\mu$m may also be a result of the stellar H$_2$O absorption. These findings highlight the need for careful interpretation of CO and H$_2$O emission at these wavelengths in disks around similarly late-type stars.

\section{Unidentified Lines} \label{sec:unidentified}
\begin{figure*}[!t]
\centering
    \includegraphics[width=0.92\textwidth, trim={0 0 0 0}, clip]{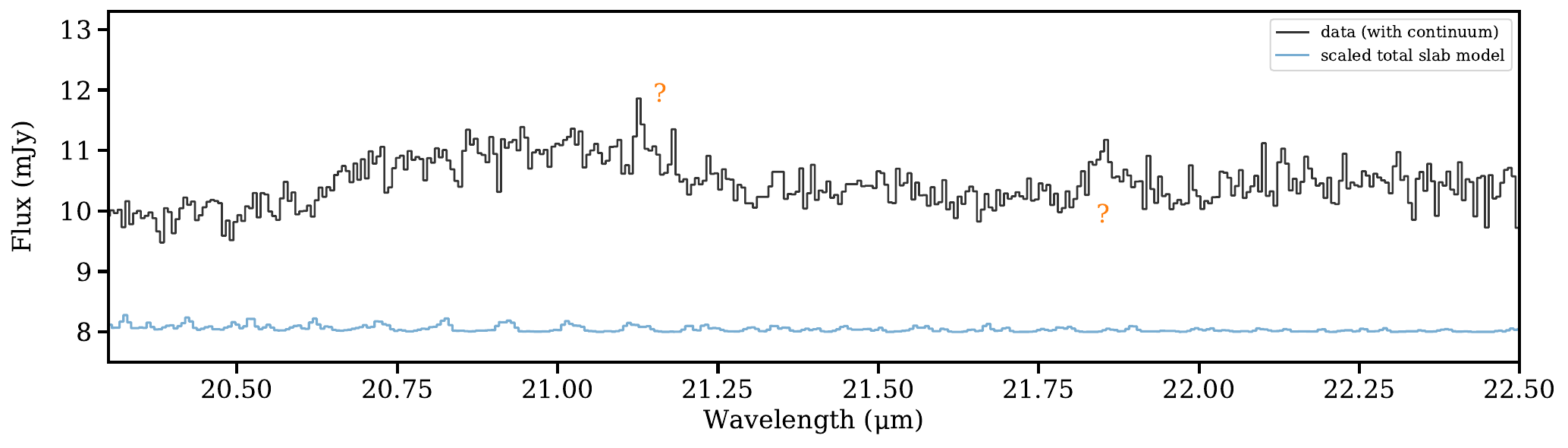}
\caption{Examples of unidentified lines (marked with ``?") in the MIRI/MRS spectrum of J0446B.   \label{fig: unidentified}}
\end{figure*}

Several line features in the spectrum of J0446B can not be identified as common species in the HITRAN database. 
A few prominent examples include the feature around 8.03\,$\mu$m that does not match CH$_4$ and H$_2$ (see Figure~\ref{fig:h2}), and the feature around 17.63\,$\mu$m that can not be explained as H$_2$O (see Figure~\ref{fig:slab-model}). Figure~\ref{fig: unidentified} also highlights the wavelength range around 20-22\,$\mu$m where multiple emission lines may present. The wide bump around 21\,$\mu$m likely corresponds to a certain dust feature that will be explored in a future analysis.

\bibliography{sample631}{}
\bibliographystyle{aasjournal}
\end{CJK*}
\end{document}